\documentclass[fleqn,twoside]{article}%
\usepackage{amsmath}
\usepackage{amsfonts}
\usepackage{amssymb}
\usepackage{graphicx}%
\usepackage{caption}
\def\be{\begin{equation}}

\def\ee{\end{equation}}
\def\bes{\begin{equation}\begin{split}&}
\def\es{\end{split}}
\def\bi{\bibitem}

\begin{document}
\title{Early universe in view of a modified theory of gravity}
\author{Ranajit Mandal\footnote{E-mail:ranajitmandalphys@gmail.com}, Dalia Saha \footnote{E-mail:daliasahamandal1983@gmail.com}, Mohosin Alam \footnote{E-mail:alammohosin@gmail.com}, Abhik Kumar Sanyal \footnote{E-mail:sanyal\_ ak@yahoo.com}}

\maketitle
\noindent
\begin{center}
\noindent
${^*}$ Dept. of Physics, Rammohan College, Kolkata, West Bengal, India - 700009.\\
\noindent
$^{\dag, \S}$ Dept. of Physics, Jangipur College, Murshidabad, West Bengal, India - 742213\\
\noindent
${^\ddag}$ Dept. of Physics, Saidpur U. N. H. S., Murshidabad, West Bengal, India - 742225. \\
\end{center}

\begin{abstract}
 We study the quantum evolution of the early universe, its semi-classical analogue together with inflationary regime, in view of a generalized modified theory of gravity. The action is built by supplementing the non-minimally coupled scalar-tensor theory of gravity with scalar curvature squared term and a Gauss-Bonnet-dilatonic coupled term. It is generalized, since all the parameters are treated as arbitrary functions of the scalar field. It is interesting to explore the fact that instead of considering additional flow parameters, an effective potential serves the purpose of finding inflationary parameters. The dilaton stabilization issue appears here as a problem with reheating. Addition of a cosmological constant term alleviates the problem, and inflation is effectively driven by the vacuum energy density. Thus Gauss-Bonnet term might play a significant role in describing late-time cosmic evolution.

\end{abstract}
\noindent
\section{\bf{Introduction}}

It is well known fact that the `standard model of cosmology' based on General theory of relativity (GTR) explains a long evolution history of the universe, right from the structure formation, and the formation of CMBR (at a redshift $z \approx 3200$) upto the recent decelerated matter dominated era (at a redshift $z \approx 1$), once the seed of perturbations is assumed to exist. Nevertheless, it has already been established that gauge-invariant divergences make GTR non-renormalizable, and also that it can not quite accommodate observations in connection Sn1a data, associated with the evolution of the universe at the late-time. In connection with the last statement, we remember that the $\Lambda$CDM model, not only suffers from the fine tuning problem, but also it is only restricted upto phantom divide line,(while observation does not omit the crossing of phantom divide line) and does not reflect the necessary (for structure formation) early Friedmann-like decelerated expansion  $a \propto t^{2\over 3}$ prior to the recent accelerated expansion. These are also true for other extended varieties of GR models, viz. the quintessence and even more exotic models, with single scalar field. On the contrary, Higher-order curvature induced actions can address these issues quite comfortably \cite{I1,I2,I3,I4,I5,I6,I7,I8,I9,I10}. Thus, GTR is required to be modified at both ends, the early and the late stages of cosmic evolution. In the present manuscript, we focus our attention to the very early stage of cosmic evolution. In this context, it has been realized long back that GTR should be replaced by a viable quantum theory of gravity. However, despite serious attempts over several decades, a renormalizable quantum theory of gravity is still not at hand. Nevertheless, all attempts in this direction lead to the weak field effective actions in $4$-dimension carrying different higher-order curvature invariant terms. In the absence of a complete and viable quantum theory of gravity, it is necessarily required to test the viability of different modified actions in the context of early stage of cosmic evolution, and to get certain insights regarding the behaviour of our universe near Planck's epoch.\\

Principle candidates of a reasonable renormalized theory of gravitational action, apart from the linear sector (Einstein-Hilbert term), are curvature squared terms, viz., $R^2$ and $R_{\mu\nu}R^{\mu\nu}$ \cite{Stelle}. Note that there is no need to include $R_{\alpha\beta\mu\nu}R^{\alpha\beta\mu\nu}$, since it combines with the other two, to form the Gauss-Bonnet term, $\mathcal{G} = R^2-4R_{\mu\nu}R^{\mu\nu}+R_{\alpha\beta\mu\nu}R^{\alpha\beta\mu\nu}$, which is topologically invariant in four dimensions. However, as the theory is expanded in the perturbative series about linearized gravity, ghost degrees of freedom appear, which destroys the unitarity. In fact, in the absence of $R_{\mu\nu}R^{\mu\nu}$ term, ghosts disappear, but ultraviolet divergence reappears, rendering the theory non-renormalizable. It was therefore attempted to construct a second derivative theory with higher order curvature invariant terms for the purpose, which ended up in Lanczos-Lovelock gravity \cite{L1, L2}. Nevertheless, Lanczos-Lovelock gravity leads to second order field equations with different combinations of higher-order terms including the Gauss-Bonnet combination, which contributes only in dimensions greater than $4$. This is because, as already mentioned, Gauss-Bonnet term is topologically invariant and thus does not contribute to the field equations in four dimensions. Likewise, the other different Lanczos-Lovelock combinations behave similarly at different dimensions. However, if we just concentrate upon Gauss-Bonnet term, it is important to mention that, this combination also arises naturally as the leading order of the $\alpha'$ expansion of heterotic superstring theory, where, $\alpha'$ is the inverse string tension \cite{S1, S2, S3, S4}. Further, the low energy limit of the string theory gives rise to the dilatonic scalar field which is found to be coupled with various curvature invariant terms \cite{D1, D2}. Therefore the leading quadratic correction gives rise to Gauss-Bonnet term with a dilatonic coupling \cite{BD}. Another very important aspect of considering such a term is: it provides nothing beyond second order field equations, and therefore appears to be free from ghost degrees of freedom. \\

Despite the fact that Lanczos-Lovelock gravity shows unitary time evolution of quantum states, when expanded perturbatively about the flat Minkowski background, nonetheless, nonperturbatively the situation is miserable. In fact, canonical analysis of 5-dimensional Lovelock action under $4 + 1$ decomposition, makes the theory intrinsically nonlinear \cite{DF}. Even its linearized version is cubic rather than quadratic. As a result, diffeomorphic invariance is not manifest and standard canonical formulation of the theory is not possible \cite{We1}. Such a situation arises due to the appearance of terms quartic in velocities in the Lagrangian. Hence, the expression for velocities is multi-valued functions of momenta, resulting in the so called multiply branched Hamiltonian with cusps. This makes classical solution unpredictable, as at any time one can jump from one branch of the Hamiltonian to the other. Further, the momentum does not provide a complete set of commuting observable, which result in non-unitary time evolution of quantum states. It has also been shown that the Einstein-Gauss-Bonnet-dilatonic action suffers from the same disease of the so-called `Branched Hamiltonian', and presently, there is no standard technique to handle this issue \cite{We1, We2}. \\

Under such critical situation, it has been shown that the presence of scalar curvature squared term ($R^2$) in the action, bypasses the pathology of Branched Hamiltonian appearing in Lanczos-Lovelock action \cite{We1, We2}. Thus, in view of the above discussions, we find it worth studying the cosmic evolution of the early universe in view of the following action,

\be\label{A}
S_c=\int d^{4}x \sqrt{-g} \bigg{[}{\alpha(\phi)R}-M_P^2 \Lambda +\beta(\phi)R^2 +\gamma(\phi)\mathcal{G}-\frac{1}{2}\phi_{,\mu}\phi^{,\mu}-V(\phi)\bigg{]}.
\ee
In the above, Gauss-Bonnet combination ($\mathcal{G}=R^2-4R_{\mu\nu}R^{\mu\nu}+R_{\mu\nu\rho\sigma}R^{\mu\nu\rho\sigma}$) is coupled with a dilatonic field, in view of the string theory or an ordinary scalar field in view of the Lanczos-Lovelock combination, through an arbitrary coupling parameter $\gamma(\phi)$, while the other two coupling parameters $\alpha(\phi)$ and $\beta(\phi)$ are also arbitrary function of $\phi$, and $V(\phi)$ is an arbitrary potential function. The action \eqref{A} generalizes our earlier one appearing in \cite{We2}, in respect of considering non-minimal coupling in view of $\alpha(\phi)$ term, and taking independent coupling functions $\beta(\phi)$ and $\gamma(\phi)$ into account. An additional cosmological constant term $\Lambda$ appearing with appropriate dimension ($M_P^2 = (8\pi G)^{-1}$) generalizes the action even further. The advantage of considering such an apparently complicated action with non-minimal coupling parameters $\alpha(\phi)$, and $\beta(\phi)$ is: it encompasses different cases starting from $\alpha(\phi) = \mathrm{constant} = {M_P^2\over 2}$ and $\beta(\phi) = \mathrm{constant}$, and also the cases in which either of the two is a constant. In fact, we have shown that such different choices lead to the de-Sitter solution of the classical field equations with identical form of potential in all the cases studied. Further, it has also been shown that all the different cases lead to the same inflationary parameters. The above action could have been further generalized by incorporating Riemann squared term $(R_{\alpha\beta\gamma\delta}R^{\alpha\beta\gamma\delta})$ with an additional coupling parameter. However, it becomes extremely difficult to handle the field equations. We shall try to pose such a generalized action in future.\\

It is now generally believed that the early universe must have passed through an inflationary era, since such a regime not only solves the horizon and flatness problems but also gives birth to the seeds of perturbations required for structure formation. Thus, inflation should be considered as a scenario rather than a model. To study how the present model \eqref{A} leads to a viable inflationary regime, canonical formulation of the action under consideration should be performed as a precursor, which is the most strong mathematical tool within gravitational field, both from theoretical as well as observational aspects. Canonical analysis is presented here as a mere prologue to canonical quantization scheme.\\

In the following section, we write the general field equations and then translate them in the isotropic and homogeneous Robertson-Walker minisuperspace model. Classical de-Sitter solutions are thereby presented. In section 3, we perform canonical formulation, which is a non-trivial task for the higher order theory under consideration. Thereafter, standard canonical quantization scheme is followed to establish hermiticity of the effective Hamiltonian operator. The viability of the quantum equation is tested under an appropriate semiclassical approximation. In section 4, slow-roll approximation is performed and the results are compared with recently released data sets. Finally concluding remarks are presented in section 5.\\

\section{Action, field equations and classical solutions:}

It has recently been established \cite{Me1} that Modified Horowitz' formalism (MHF) and the Dirac constraint analysis towards canonical formulation of higher-order theory of gravity lead to the same Hamiltonian, provided Dirac algorithm is initiated only after taking care of the divergent terms appearing in the action. In fact, it has been proved that otherwise, although the two Hamiltonians are canonically related, they produce completely different quantum descriptions \cite{Me1}. We shall here follow the MHF which bypasses Dirac's algorithm even in the presence of the Lapse function $N(t)$ \cite{We1, We2, Me1, M1, M2, M3, M4, M5}, that essentially acts as a Lagrange multiplier. Dirac's technique is explored in the appendix, to exhibit equivalence. For the purpose of carrying out MHF, the action \eqref{A} is required to be supplemented by appropriate boundary terms, viz.,

\be\label{Scf}
S_c=\int d^{4}x \sqrt{-g} \bigg{[}{\alpha(\phi)R}-\Lambda M^2_P+\beta(\phi)R^2+\gamma(\phi)\mathcal{G}-\frac{1}{2}\phi_{,\mu}\phi^{,\mu}-V(\phi)\bigg{]}+ \alpha(\phi)\Sigma_{R}+\beta(\phi)\Sigma_{R^2}+\gamma(\phi)\Sigma_{\mathcal{G}}. \ee
In the above, $\alpha(\phi)\Sigma_{R}=2\alpha(\phi)\oint_{\partial\nu}K\sqrt{h}d^3x;~ \beta(\phi)\Sigma_{R^2}= 4\beta(\phi)\oint_{\partial\nu}RK\sqrt{h}d^3x = \beta(\phi)(\Sigma_{R_1^2}+\Sigma_{R_2^2})=4\beta(\phi)\oint_{\partial\nu}[ {^3R}+ ({^4R}-{^3R})]RK\sqrt{h}d^3x; ~ \gamma(\phi)\Sigma_{\mathcal{G}} = 4 \gamma(\phi)\oint_{\partial\nu}\big{(}2G_{ij}K^{ij}+\frac{\textsf{K}}{3}\big{)}\sqrt{h}d^3x$ are the supplementary boundary terms which are modified versions of Gibbons-Hawkings-York term \cite{GHY, GHY1, GHY2}, while $\textsf{K}=K^2-3KK^{ij}K_{ij}+2K^{ij}K_{ik}K^k_j$, $K_{ij}$ being the extrinsic curvature tensor, $h$ is the determinant of the induced metric $h_{ij}$, and $K = K^{ij} h_{ij}$. Now, the field equation is found under the standard metric variation as \cite{fld},

\be\label{gfield}\begin{split} 2\Big(\alpha(\phi)G_{\mu\nu}+\Box\alpha(\phi) g_{\mu\nu} -\nabla_{\mu}\nabla_{\nu}\alpha(\phi)\Big)&+ 4\Big[\beta(\phi)RR_{\mu\nu} - g_{\mu\nu} \Box(\beta(\phi)R)-\nabla_{\mu}\nabla_{\nu}(\beta(\phi) R)
-{1\over 8}g_{\mu\nu}\beta(\phi)R^2\Big] \\&+2\gamma(\phi)H_{\mu\nu}+8(\gamma''\nabla^{\rho}\phi\nabla^{\sigma}\phi + \gamma'\nabla^{\rho}\nabla^{\sigma}\phi)P_{\mu\rho\nu\sigma}-\Lambda M^2_P={T^{(\phi)}_{\mu\nu}},\end{split}\ee
where, $G_{\mu\nu}=R_{\mu\nu}-{1\over 2}g_{\mu\nu}R$ and $T^{(\phi)}_{\mu\nu}=\nabla_{\mu}\phi\nabla_{\nu}\phi-{1\over 2}g_{\mu\nu}\nabla_{\lambda}\phi\nabla^{\lambda} \phi-g_{\mu\nu}V(\phi)$ are the Einstein tensor and the energy-momentum tensor respectively. Further, $H_{\mu\nu}=2\big(RR_{\mu\nu}-2R_{\mu\rho}R^{\rho}_{\nu}-2R_{\mu\rho\nu\sigma}R^{\rho\sigma}+R_{\mu\rho\sigma\lambda}R^{\sigma\rho\lambda}-{1\over 2}g_{\mu\nu} \mathcal{G}\big)$ and $P_{\mu\nu\rho\sigma}=R_{\mu\nu\rho\sigma}+2g_{\mu[\sigma} R_{\rho]\nu}+2g_{\nu[\rho} R_{\sigma]\mu}+Rg_{\mu[\rho} g_{\sigma]\nu}$. Explicit form of equation \eqref{gfield} together with  the $\phi$ variation equation are expressed as,

\be \label{phivar1}\begin{split}& \alpha G_{\mu\nu} + \Box\alpha g_{\mu\nu} - \alpha_{;\mu;\nu} + \Big[2\beta\big(RR_{\mu\nu} - {1\over 4}g_{\mu\nu}R^2\big) - 2\big(\Box{(\beta R)} g_{\mu\nu} + (\beta R)_{;\mu;\nu}\big)\Big]\\&
\hspace{0.9 cm}+ 2\gamma \Big[RR_{\mu\nu} - 2 R_{\mu\rho}R^{\rho}_{\nu} - 2R_{\mu\rho\nu\sigma}R^{\rho\sigma} + R_{\mu\rho\sigma\lambda}R_\nu^{\sigma\rho\lambda} - {1\over 4}g_{\mu\nu}\left(R^2-4R_{\mu \nu} R^{\mu \nu}+R_{\mu \nu \delta \gamma} R^{\mu \nu \delta \gamma}\right)\Big]\\&
\hspace{0.9 cm}+4\left(\gamma''\phi^{;\rho}\phi^{;\sigma}+\gamma'\phi^{;\rho;\sigma}\right)\big[R_{\mu\rho\nu\sigma} + 2g_{\mu[\sigma}R_{\rho]\nu} + 2g_{\nu[\rho}R_{\sigma]\mu} + R g_{\mu[\rho}g_{\sigma]\nu} \big]-{\Lambda M^2_P\over 2} = {T_{\mu\nu}\over 2}\\&
\Box\phi-\alpha' R-\beta' R^2-\gamma' \mathcal{G}-V'=0, \end{split}\ee
where, prime denotes derivative with respect to $\phi$. In the homogeneous and isotropic Robertson-Walker metric, viz.,

\be \label{RW} ds^2 = -N^2(t)dt^2+a^2(t)\Big[\frac{dr^2}{1-kr^2}+r^2(d\theta^2+sin^2\theta\; d\phi^2)\Big],
\ee
the Ricci scalar as well as the Gauss-Bonnet terms read as,

\be \label{R} {R}=\frac{6}{N^2}\Big(\frac{\ddot{a}}{a}+\frac{\dot{a}^{2}}{a^2}+N^2\frac{k}{a^{2}}-\frac{\dot a\dot N}{aN}\Big);\hspace{1.5 cm} \mathcal{G} = {24\over N^3 a^3}(N \ddot a - \dot N\dot a)\left({\dot a^2\over N^2} + k\right).\ee
The $a$ variation i.e. ($^i_i$) equation, the $N$ variation i.e. the $(^0_0)$ equation and the $\phi$ variation equation are as follows,

\be \label{avariation} \begin{split}&2\alpha\bigg[\frac{2\ddot{a}}{a}+\frac{\dot{a}^2}{a^2}+\frac{k}{a^2}\bigg]+2\alpha'\bigg[\ddot{\phi}+\frac{2\dot{a}\dot{\phi}}{a}\bigg]
+2\alpha''\dot{\phi}^2+12\beta\bigg[\frac{2\ddddot{a}}{a}+\frac{4\dot{a}\dddot{a}}{a^2}+\frac{3\ddot{a}^2}{a^2}-\frac{12\dot{a}^2\ddot{a}}{a^3}
+\frac{3\dot{a}^4}{a^4}-\frac{4k\ddot{a}}{a^3}+\frac{2k\dot{a}^2}{a^4}-\frac{k^2}{a^4}\bigg]\\& \hspace{1.2 cm}+48\beta'\dot{\phi}\bigg[\frac{\dddot{a}}{a}+\frac{2\dot{a}\ddot{a}}{a^2}-\frac{\dot{a}^3}{a^3}-\frac{k\dot{a}}{a^3}\bigg]+(24\beta''\dot{\phi}^2+24\beta'\ddot{\phi})\bigg[\frac{\ddot{a}}{a}+\frac{\dot{a}^2}{a^2}+\frac{k}{a^2}\bigg]
+\frac{16\gamma'\ddot{a}\dot{a}\dot{\phi}}{a^2}+8\gamma'\ddot{\phi}\bigg[\frac{\dot{a}^2}{a^2}+\frac{k}{a^2}\bigg]\\&\hspace{6.1 cm}+8\gamma''\dot{\phi}^2\bigg[\frac{\dot{a}^2}{a^2}+\frac{k}{a^2}\bigg]+\frac{\dot{\phi}^2}{2}-V-\Lambda M^2_P=0.\end{split}\ee

\be \label{00} \begin{split}6\alpha\bigg{(}{\dot a^2\over a^2}+ {k \over a^2}\bigg{)} + 6\alpha'\dot\phi\Big({\dot a\over a}\Big) &+36\beta \bigg{(}\frac{2\dot a\dddot a}{a^2}-\frac{\ddot a^2}{a^2} +\frac{2\dot a^2 \ddot a}{a^3}-\frac{3\dot a^4}{a^4}-{2k\dot a^2\over a^4}+{k^2\over a^4}\bigg{)}\\&+72\beta'\dot\phi\bigg{(}\frac{\dot a\ddot a}{a^2}+\frac{\dot a^3}{a^3}+{k\dot a\over a^3}\bigg{)}+24\gamma'\dot\phi\bigg{(}{\dot a^3\over a^3}+{k\dot a\over a^3}\bigg{)}-\Lambda M^2_P= \bigg{(}\frac{\dot\phi^2}{2}+V \bigg{)}, \end{split} \ee

\be \label{phivariation} \begin{split}& \ddot\phi +3{\dot a\over a} \dot\phi + V' - 6\alpha'\Big({\ddot a\over a}+{\dot a^2\over a^2} + {k\over a^2}\Big)- 36\beta'\Big({\ddot a^2\over a^2}+2{\dot a^2\ddot a\over a^3}+ \frac{\dot a^4}{a^4}+2{k\ddot a\over a^3}+{2k\dot a^2\over a^4}+{k^2\over a^4}\Big) -24\gamma'\Big({\dot a^2\ddot a\over a^3}+ {k\ddot a\over a^3}\Big)  =0.\end{split}\ee
Not all the above components of Einstein's equations are independent, e.g. the $(^0_0)$ equation is the energy constraint equation. Thus it is suffice to consider only the two independent components of Einstein's equations viz. (\ref{00}) and (\ref{phivariation}), for all practical purpose. A viable gravity theory must admit de-Sitter solution in vacuum. It is not difficult to see that the above field equations also admit the following de-Sitter solutions,

\be\label{aphi} a = a_0e^{\lambda t}; ~~~ \phi= \phi_0e^{-\lambda t},\ee
in the spatially flat space ($k=0$), under the following different conditions,\\

\noindent
\textbf{Case 1:}\\
\be\label{param}\alpha(\phi) = {\alpha_0\over \phi};~~ V(\phi) = \frac{1}{2} \lambda^2 \phi^2-\Lambda M^2_P;~~\mathrm{and}~~ 6\beta + \gamma = - {1\over 2\lambda^2}\left({\alpha_0\over \phi} +{\phi^2\over 24}\right),\ee
where, $a_0$, $\phi_0$, $\alpha_0$ and $\lambda$ are arbitrary constants while $\beta(\phi)$ and $\gamma(\phi)$ retain arbitrary forms of $\phi$, being related as above after setting the constant of integration to zero, without any loss of generality. Since we shall encounter operator ordering ambiguities during canonical quantization, therefore we remove the arbitrariness on $\beta(\phi)$ and $\gamma(\phi)$, following a simple assumption viz,

\noindent
\textbf{Subcase 1a:}\\
\be\label{betagamma} \beta = -\frac{\alpha_0}{12\lambda^2\phi} = {\beta_0\over \phi};~~~\gamma = -\frac{\phi^2}{48\lambda^2} = {\gamma_0\phi^2}\hspace{0.5 cm} \mathrm{where}, \hspace{0.5 cm} \beta_0 = -{\alpha_0\over 12 \lambda^2};~~~\gamma_0 = -{1\over 48\lambda^2}.\ee
The above forms of $\alpha(\phi)$ \eqref{param}, $\beta(\phi)$ and $\gamma(\phi)$ \eqref{betagamma} will be used during canonical quantization (for operator ordering, in particular), semiclassical approximation, as well as during slow roll approximation.\\
\textbf{Subcase 1b:}\\
\be\label{betagamma1b} \beta =-\frac{\phi^2}{288\lambda^2} = {\beta_0 \phi^2};~~~\gamma =-\frac{\alpha_0}{2\lambda^2\phi}  = {\gamma_0\over \phi}\hspace{0.5 cm} \mathrm{where}, \hspace{0.5 cm} \beta_0 = -{1\over 288\lambda^2};~~~\gamma_0 = -{\alpha_0\over 2 \lambda^2} .\ee

\noindent
\textbf{Case 2:}\\
If $\beta=\beta_0 =$ constant, then the above de-Sitter solution \eqref{aphi} is admissible under the same choice of the potential, under the condition,

\be\label{betagamma2}\alpha(\phi) = {\alpha_0\over \phi};~~ V(\phi) = \frac{1}{2} \lambda^2 \phi^2-\Lambda M^2_P;~~\mathrm{and} ~~\gamma=- {1\over 2\lambda^2}\left({\alpha_0\over \phi} +{\phi^2\over 24}\right).\ee

\noindent
\textbf{Case 3:}\\
If $\alpha= \alpha_0 =$ constant, even then the same de-Sitter solution is admissible under the condition,
\be\label{betagamma3} V(\phi) = \frac{1}{2} \lambda^2 \phi^2-\Lambda M^2_P+6\lambda^2\alpha_0;~~\mathrm{and}~~~ 6\beta + \gamma = - \left({\phi^2\over 48\lambda^2}\right).\ee
It is to be noted that the additional term appearing in the potential being a constant may be absorbed in the potential, keeping its form unaltered, as well.\\

\noindent
\textbf{Case 4:}\\
If $\alpha=\alpha_0 =$constant,~~and ~ $\beta= \beta_0 =$ constant, ~~then,
\be\label{betagamma4} V(\phi) = \frac{1}{2} \lambda^2 \phi^2-\Lambda M^2_P+6\lambda^2\alpha_0;~~\mathrm{and}~~ \gamma=-\frac{\phi^2}{48\lambda^2}.\ee
Here again the form of the potential only differs from the earlier ones only by a constant term as in case 3.

\section{Canonical formulation:}

In this section, our aim is to establish the canonical structure of action (\ref{Scf}) in the Robertson-Walker minisuperspace model \eqref{RW}. This is possible once we can write it in terms of the basic variables $h_{ij} = a^2\delta_{ij} = z\delta_{ij}$, and $k_{ij} = -\frac{\dot h_{ij}}{2N}=-\frac{a\dot a}{N}\delta_{ij}=\frac{\dot z}{2N}\delta_{ij}$, where, $a^2 = z$. For this purpose, we first express the action (\ref{Scf}) in terms of $h_{ij} = a^2\delta_{ij} = z\delta_{ij}$ using the form of the Ricci scalar and the Gauss-Bonnet term \eqref{R} as,

\begin{center}\be \label{Sc1}\begin{split} S_c = \int \bigg{[}&{3\alpha(\phi)\sqrt{z}\bigg{(}\frac{\ddot z}{N}-\frac{\dot z\dot N}{N^2}+2kN\bigg{)}}-Nz^{3\over 2}\Lambda M^2_P+\frac{9\beta(\phi)}{\sqrt z}\bigg{(}\frac{\ddot z^2}{N^3}-\frac{2\dot z \ddot z \dot N}{N^4}+\frac{\dot z^2\dot N^2}{N^5}-\frac{4k\dot z\dot N}{N^2}+\frac{4k\ddot z}{N}+4k^2N \bigg{)}\\&+\frac{3\gamma(\phi)}{N\sqrt z}\bigg{(}\frac{\dot z^2\ddot z}{N^2z}-\frac{\dot z^4}{2N^2z^2}-\frac{\dot z^3\dot N}{N^3z}+4k\ddot z-\frac{2k\dot z^2}{z}-\frac{4k\dot z\dot N}{N} \bigg{)}+z^{\frac{3}{2}}\bigg{(}\frac{1}{2N}\dot{\phi}^2-V(\phi)N\bigg{)}\bigg{]}dt \\& \hspace{5.10 cm}+ \alpha(\phi)\Sigma_{R}+ \beta(\phi)\Sigma_{R^2}+\gamma(\phi) \Sigma_{\mathcal{G}}. \end{split}\ee\end{center}
In the above, $\alpha(\phi)\Sigma_{R}=- \alpha(\phi)\frac{3\sqrt z \dot z}{N}$, $\beta(\phi)\Sigma_{R_1^2}=- \beta(\phi)\frac{36 k \dot z}{N\sqrt z}$, $\beta(\phi)\Sigma_{R_2^2}=- \beta(\phi)\frac{18\dot z}{N^3\sqrt z}\textbf{(}\ddot z-\frac{\dot z\dot N}{N}\textbf{)}$ and $\gamma(\phi)\Sigma_{\mathcal{G}}=- \gamma(\phi)\frac{\dot z}{N\sqrt z}\textbf{(}\frac{\dot z^2}{N^2 z}+12k\textbf{)}$ are the supplementary surface terms as already mentioned, but now in the isotropic and homogeneous Robertson-Walker metric \eqref{RW}. It is important to mention that unlike GTR, here the lapse function appears in the action with its time derivative, behaving like a true variable. Despite such uncanny situation, one can still bypass Dirac's algorithm as we demonstrate below.\\

First, under integrating the above action \eqref{Sc1} by parts, the counter terms $\alpha(\phi)\Sigma_R$, $\beta(\phi)\Sigma_{R_1^2}$ and $\gamma(\phi)\Sigma_{\mathcal{G}}$ get cancelled with the divergent (total derivative) terms and we arrive at,

\be \label{Sc2}\begin{split} S_c&=\int \bigg{[}{\bigg{(}-\frac{3\alpha'\dot\phi\dot z\sqrt z}{N}-\frac{3\alpha\dot z^2}{2N\sqrt z}+6kN\alpha\sqrt z\bigg{)}}-Nz^{3\over 2}\Lambda M^2_P+\frac{9\beta}{\sqrt z}\bigg{(}\frac{\ddot z^2}{N^3}-\frac{2\dot z \ddot z \dot N}{N^4}+\frac{\dot z^2\dot N^2}{N^5}+\frac{2k{\dot z}^2}{Nz}+4k^2N \bigg{)}\\& \hspace{1.30 cm}-\frac{36\beta' k \dot z \dot\phi}{N\sqrt z}-\frac{\gamma'\dot z \dot\phi}{N\sqrt z}\bigg{(}\frac{\dot z^2}{N^2z}+12k \bigg{)}+z^{\frac{3}{2}}\bigg{(}\frac{1}{2N}\dot{\phi}^2-VN\bigg{)}\bigg{]}dt +\beta(\phi)\Sigma_{R_2^2}. \end{split}\ee
At this stage we construct an auxiliary variable $Q$ under the following prescription,

\be
\label{Q} Q = \frac{\partial S_c}{\partial\ddot z}=\frac{18\beta}{N^3\sqrt z}\bigg{(}\ddot z-\frac{\dot N\dot z}{N}\bigg{)},
\ee
and substitute it judiciously straight into the action (\ref{Sc2}), as a result of which it now reads as,

\be \label{Sc3} \begin{split} S_c&=\int \bigg{[} {\bigg{(}}-\frac{3\alpha'\dot\phi \dot z\sqrt z}{N}-\frac{3\alpha \dot z^2}{2N\sqrt z}+6kN\alpha \sqrt z \bigg{)}-Nz^{3\over 2}\Lambda M^2_P+Q\ddot z-\frac{N^3\sqrt z Q^2}{36\beta}-\frac{\dot N\dot z Q}{N}+\frac{18\beta k \dot z^2}{Nz^{\frac{3}{2}}}+\frac{36\beta N k^2}{\sqrt z}\\& \hspace{1.30 cm}-\frac{36\beta' k\dot \phi \dot z}{N\sqrt z}-\frac{\gamma'\dot z \dot\phi}{N\sqrt z}\bigg{(}\frac{\dot z^2}{N^2z}+12k \bigg{)}+z^{\frac{3}{2}}\bigg{(}\frac{1}{2N}\dot{\phi}^2-VN\bigg{)}\bigg{]}dt + \beta(\phi)\Sigma_{R_2^2}.\end{split}\ee
Now, further integrating \eqref{Sc3} by parts, the rest of the surface terms viz. $\beta(\phi)\Sigma_{R_2^2}$ gets cancelled with the total derivative term yet again, and thus the action (\ref{Sc1}) being free of all the divergent terms appearing under variation is finally expressed as,

\be \label{Sc44} \begin{split} S_c&=\int \bigg{[} {\bigg{(}}-\frac{3\alpha'\dot\phi \dot z\sqrt z}{N}-\frac{3\alpha \dot z^2}{2N\sqrt z}+6kN\alpha \sqrt z \bigg{)}-Nz^{3\over 2}\Lambda M^2_P-\dot Q\dot z-\frac{N^3\sqrt z Q^2}{36\beta}-\frac{\dot N\dot z Q}{N}+\frac{18\beta k \dot z^2}{Nz^{\frac{3}{2}}}+\frac{36\beta N k^2}{\sqrt z}\\&\hspace{1.30 cm} -\frac{36\beta' k\dot \phi \dot z}{N\sqrt z}-\frac{\gamma'\dot z \dot\phi}{N\sqrt z}\bigg{(}\frac{\dot z^2}{N^2z}+12k \bigg{)}+z^{\frac{3}{2}}\bigg{(}\frac{1}{2N}\dot{\phi}^2-VN\bigg{)}\bigg{]}dt.\end{split}\ee
The canonical momenta now read as,

\begin{eqnarray}
  p_Q &=& -\dot z, \label{pQ}\\
  p_z &=& -\frac{3\alpha'\dot\phi\sqrt z}{N}-\frac{3\alpha \dot z}{N\sqrt z}-\dot Q-\frac{\dot N Q}{N}+\frac{36\beta k\dot z}{Nz^{\frac{3}{2}}}-\frac{36\beta' k \dot\phi}{N\sqrt z} -\frac{3\gamma'\dot\phi}{N\sqrt z}\bigg{(}\frac{\dot z^2}{N^2 z}+4k\bigg{)},\label{pz}\\
  p_{\phi} &=& -\frac{3\alpha'\dot z\sqrt z}{N}-\frac{36\beta' k \dot z}{N\sqrt z}-\frac{\gamma'\dot z}{N\sqrt z}\bigg{(}\frac{\dot z^2}{N^2 z}+12k\bigg{)}+\frac{z^{\frac{3}{2}} \dot\phi}{N}, \label{pph}\\
  p_N &=& -\frac{Q\dot z}{N},
\end{eqnarray}
and thus one can now, in principle, construct the phase-space structure of the Hamiltonian. However, the above relations signal the present of a momentum constraint in the form $Qp_Q + Np_N =0$. Nevertheless, one can still avoid Dirac's algorithm by constructing the following relation in view of the above momenta as,
\begin{center}
   \be\label{PN} p_Q p_z=\frac{3\alpha' \dot z \dot\phi \sqrt z}{N}+\frac{3\alpha \dot z^2}{N\sqrt z}+\dot z\dot Q+\frac{\dot N\dot z Q}{N}-\frac{36k\beta \dot z^2}{Nz^{\frac{3}{2}}}+\frac{36\beta' k\dot z \dot\phi}{N\sqrt z}+\frac{3\gamma'\dot z  \dot\phi}{N\sqrt z}\bigg{(}\frac{\dot z^2}{N^2 z}+4k\bigg{)}.\ee
\end{center}
Using the above relation \eqref{PN} and the definitions of momenta $p_Q, p_z, p_{\phi}$, the phase-space structure of the Hamiltonian constraint equation may be expressed as \footnote{Instead of $Q$ if the auxiliary variable would have been chosen as $q = NQ$, then $\dot N$ would not have appeared in the action, and essentially it would have been free from the constraint mentioned above. This is mentioned just to appreciate the fact that the complication appears here is an artefact of a bad choice of co-ordinate.},

\be \label{Sc4} \begin{split} H_c = 3\alpha \bigg{(}&\frac{{p_Q^2}}{2N\sqrt z}-2kN\sqrt z\bigg{)}-\frac{3\alpha' p_Q p_{\phi}}{z}+\frac{9\alpha'^2 p_Q^2}{2N\sqrt z}-p_Q p_z + \frac{N^3Q^2\sqrt z}{36\beta}-\frac{18k\beta}{\sqrt z}\bigg{(}\frac{{p_Q^2}}{Nz}+2kN\bigg{)}+\frac{\gamma'^2 {p_Q^6}}{2N^5z^{\frac{9}{2}}}\\&+ \frac{648k^2\beta'^2{p_Q^2}}{Nz^{\frac{5}{2}}} + \frac{72k^2\gamma'^2 p_Q^2}{Nz^{\frac{5}{2}}} -\frac{36k\beta'{p_Q} {p_{\phi}}}{z^2}+\frac{12k\gamma'^2 {p_Q^4}}{N^3z^{\frac{7}{2}}}-\frac{\gamma' {{p_Q^3}}{p_{\phi}}}{N^2z^3} -\frac{12k\gamma' p_Q p_{\phi}}{z^2} +\frac{36k\beta'\gamma' p_Q^4}{N^3z^{\frac{7}{2}}}\\& +\frac{432k^2\beta'\gamma' p_Q^2}{Nz^{\frac{5}{2}}}+\frac{108k\alpha'\beta' p_{Q}^2}{Nz^{\frac{3}{2}}} +\frac{3\alpha' \gamma' p_Q^4}{N^3 z^{\frac{5}{2}}}+\frac{36k\alpha'\gamma' p_Q^2}{Nz^{\frac{3}{2}}}+\frac{Np_{\phi}^2}{2z^{\frac{3}{2}}} + N V z^{\frac{3}{2}} +Nz^{3\over 2}\Lambda M^2_P=0. \end{split}\ee
The phase-space structure of the Hamiltonian for the higher order theory under consideration has thus been framed in (\ref{Sc4}). However, note that diffeomorphic invariance has not yet been established. Further, the Hamiltonian contains momentum $p_Q$ upto sixth order, which does not allow to obtain a viable quantum counterpart under standard canonical quantization scheme, since, this is not at all convenient in order to form the operators. Even if one does, a large number of initial (boundary) conditions are required to solve the quantum equation, which are not available. This envisages the fundamental importance of basic variable required for quantization, which are $(h_{ij}, K_{ij})$. Therefore, we need to make a canonical transformation from $\{Q, p_Q\}$ to $\{x, p_x \}$, where, $N x= \dot z$, which is performed following the prescription $Q=\frac{p_x}{N}$ and $p_Q = -Nx$. Therefore, the phase-space structure of the Hamiltonian can be expressed in terms of basic variables as,

\be \label{Hc} \begin{split} H_c&= N\bigg{[}xp_z+3\alpha \bigg{(}\frac{x^2}{2\sqrt z}-2k\sqrt z\bigg{)}+\frac{3\alpha' x p_{\phi}}{z}+\frac{9\alpha'^2 x^2}{2\sqrt z} + \frac{\sqrt z p_x^2}{36\beta}-\frac{18k\beta}{\sqrt z}\bigg{(}\frac{x^2}{z}+2k\bigg{)}+\frac{\gamma'^2 {x}^6}{2z^{\frac{9}{2}}}\\&+\frac{648k^2\beta'^2{x}^2}{z^{\frac{5}{2}}} + \frac{72k^2\gamma'^2 x^2}{z^{\frac{5}{2}}} +\frac{36k\beta'{x} {p_{\phi}}}{z^2}+\frac{12k\gamma'^2 {x}^4}{z^{\frac{7}{2}}}+\frac{\gamma' {{x}^3}{p_{\phi}}}{z^3} +\frac{12k\gamma' x p_{\phi}}{z^2} +\frac{36k\beta'\gamma' x^4}{z^{\frac{7}{2}}}\\& +\frac{432k^2\beta'\gamma' x^2}{z^{\frac{5}{2}}} +\frac{p_{\phi}^2}{2z^{\frac{3}{2}}}+\frac{108k\alpha'\beta' x^2}{z^{\frac{3}{2}}} +\frac{3\alpha' \gamma' x^4}{z^{\frac{5}{2}}}+\frac{36k\alpha'\gamma' x^2}{z^{\frac{3}{2}}} +Vz^{\frac{3}{2}}+z^{3\over 2}\Lambda M^2_P\bigg{]}=N\mathcal{H}=0, \end{split}\ee
and diffeomorphic invariance is established. The action (\ref{Sc3}) can now also be expressed in the canonical form with respect to the basic variables as,

\be \label{CF} S_c=\int\bigg{(}\dot zp_z+\dot xp_x+\dot\phi p_{\phi}-N\mathcal{H}\bigg{)}dt d^3x=\int\bigg{(}\dot{h}_{ij}\pi^{ij}+\dot{K}_{ij}{\Pi}^{ij}+\dot\phi p_{\phi}-N\mathcal{H}\bigg{)}dt d^3x,\ee
where $\pi^{ij}$ and $\Pi^{ij}$ are momenta canonically conjugate to $h_{ij}$ and $K_{ij}$ respectively. Thus the very importance of the use of basic variables have also been established. In appendix B, we construct Hamilton's equations and hence compute the field equations to demonstrate that the above the Hamiltonian \eqref{Hc} gives the correct description of the theory \eqref{A} under consideration in the minisuperspace \eqref{RW}.

\subsection{Canonical quantization:}

The quantum counterpart of the Hamiltonian (\ref{Hc}) under standard canonical quantization reads as,

\be \label{qh} \begin{split} \frac{i\hbar}{\sqrt z}\frac{\partial\Psi}{\partial z}&=-\frac{\hbar^2}{36\beta x}\bigg{(}\frac{\partial^2}{\partial x^2} +\frac{n}{x}\frac{\partial}{\partial x}\bigg{)}\Psi-\frac{\hbar^2}{2xz^2}\frac{\partial^2\Psi}{\partial \phi^2}+\frac{3}{z^{\frac{3}{2}}}\widehat{\alpha' }\widehat{p_{\phi}}\Psi+\frac{36k}{z^{\frac{5}{2}}}\widehat{\beta'}\widehat{p_{\phi}}\Psi+\bigg{(}\frac{x^2}{z^{\frac{7}{2}}}+\frac{12k}{z^{\frac{5}{2}}} \bigg{)}\widehat{\gamma'}\widehat{p_{\phi}}\Psi+\frac{9x}{2z}\widehat{\alpha'^2}\Psi\\&+\frac{648k^2x}{z^3}\widehat{\beta'^2}\Psi +\bigg{(} \frac{x^5}{2z^5} +\frac{12kx^3}{z^4} +\frac{72k^2x}{z^3}\bigg{)}\widehat{\gamma'^2}\Psi+\frac{108kx}{z^2}\widehat{\alpha'}\widehat{\beta'}\Psi +\bigg{(}\frac{3x^3}{z^3} +\frac{36kx}{z^2}\bigg{)}\widehat{\alpha'}\widehat{\gamma'}\Psi \\&+\bigg{(}\frac{36kx^3}{z^4}+ \frac{432k^2x}{z^3}\bigg{)}\widehat{\beta'}\widehat{\gamma'} \Psi +\bigg{(}\frac{3\alpha x}{2z}-\frac{6k\alpha}{x} -\frac{18kx\beta}{z^2}-\frac{36k^2\beta}{xz}+\frac{z}{x}\left(V+\Lambda M^2_P\right)\bigg{)}\Psi, \end{split}\ee
where, $n$ is the operator ordering index which removes some but not all of the operator ordering ambiguities appearing between $\hat x$ and $\hat{p_x}$. One can take note of the fact that the above equation is still having some operator ordering ambiguities, which may only be removed knowing explicit functional forms of $\alpha(\phi)$, $\beta(\phi)$ and $\gamma(\phi)$. As committed, we use the form of $\alpha(\phi)$, $\beta(\phi)$ and $\gamma(\phi)$ appearing in the classical de-Sitter solution \eqref{aphi}, \eqref{param} and \eqref{betagamma}, to perform Weyl symmetric ordering carefully between $\hat \alpha'$ and $\hat p_{\phi}$, $\hat \beta'$ and $\hat p_{\phi}$ and $\hat \gamma'$ and $\hat p_{\phi}$. As a result, the equation (\ref{qh}) now takes the form,

\be \label{qh1} \begin{split} \frac{i\hbar}{\sqrt z}\frac{\partial\Psi}{\partial z}=\Bigg[-\frac{\hbar^2 \phi}{36\beta_0 x}\left(\frac{\partial^2}{\partial x^2} +\frac{n}{x}\frac{\partial}{\partial x}\right)&-\frac{\hbar^2}{2xz^2}\frac{\partial^2}{\partial \phi^2}+\frac{3i\hbar \alpha_0}{z^{\frac{3}{2}}}\left(\frac{1}{\phi^2}\frac{\partial}{\partial \phi}-\frac{1}{\phi^3}\right) -\frac{i\hbar{\gamma_0}x^2}{z^{\frac{7}{2}}} \left(2\phi\frac{\partial}{\partial\phi}+1\right)\\& +\left(\frac{9\alpha_0^2 x}{2z\phi^4} -{6\alpha_0\gamma_0 x^3\over z^3\phi} + \frac{2\gamma_0^2 x^5\phi^2}{z^5}+ \frac{3 \alpha_0 x}{2z\phi}+{\lambda^2 z\phi^2\over 2x}+{\Lambda M^2_P z\over x}\right)\Bigg]\Psi. \end{split}\ee
We have expressed the quantum equation \eqref{qh1} in spatially flat space $k = 0$, to avoid unnecessary complication. Now, under a change of variable, the above modified Wheeler-de-Witt equation, takes the look of Schr\"{o}dinger equation, viz.,

\be \label{qh2} \begin{split} {i\hbar}\frac{\partial\Psi}{\partial \sigma}&=\left[-\frac{\hbar^2 \phi}{54\beta_0 x}\left(\frac{\partial^2}{\partial x^2} +\frac{n}{x}\frac{\partial}{\partial x}\right) -\frac{\hbar^2}{3x\sigma^{\frac{4}{3}}}\frac{\partial^2}{\partial \phi^2}+\frac{2i\hbar \alpha_0} {\sigma}\left(\frac{1}{\phi^2}\frac{\partial}{\partial \phi}-\frac{1}{\phi^3} \right) - \frac{2i\hbar \gamma_0 x^2}{3\sigma^{7\over 3}}\left(2\phi{\partial\over \partial \phi} + 1\right) +V_e\right]\Psi\\&=\widehat H_e \Psi, \end{split}\ee
where, the proper volume, $\sigma=z^{\frac{3}{2}} = a^3$ plays the role of internal time parameter. In the above equation, the effective potential $V_e$ is given by,
\begin{center}
    \be V_e = \frac{3\alpha_0^2 x}{\sigma^{2\over 3}\phi^4} - \frac{4\alpha_0\gamma_0 x^3}{\sigma^2 \phi} + \frac{4\gamma_0^2 x^5\phi^2}{3\sigma^{10\over 3}} + \frac{\alpha_0 x}{\sigma^{2\over 3}\phi} + \frac{\lambda^2\sigma^{2\over 3}\phi^2}{3x}+{2\sigma^{2\over 3}\Lambda M^2_P\over x}. \ee
\end{center}

\subsection{Hermiticy of $\widehat{H_e}$ and probabilistic interpretation:}

We split the effective Hamiltonian operator $\widehat H_e$ (\ref{qh2}) into three components apart from the effective potential as,

\be\label{He}\begin{split} \widehat H_e= &-\frac{\hbar^2\phi}{54\beta_0 x}\bigg{(}\frac{\partial^2}{\partial x^2} +\frac{n}{x}\frac{\partial}{\partial x}\bigg{)}-\frac{\hbar^2}{3x\sigma^{\frac{4}{3}}}\frac{\partial^2}{\partial \phi^2}+\frac{2i\hbar } {3\sigma}\bigg{(}\frac{3\alpha_0}{\phi^2}-\frac{2{\gamma_0}\phi x^2}{\sigma^{\frac{4}{3}}} \bigg{)} \frac{\partial}{\partial \phi} +\frac{2i\hbar}{\sigma} \bigg{(}\frac{\gamma_0 x^2}{3 \sigma^\frac{4}{3}}-{\alpha_0\over\phi^3}\bigg{)}+V_e\\&=\widehat H_1+\widehat H_2+\widehat H_3+\widehat V_e,\end{split}\ee
where,

\begin{eqnarray}
\widehat H_1 &=& -\frac{\hbar^2\phi}{54\beta_0 x}\bigg{(}\frac{\partial^2}{\partial x^2} +\frac{n}{x}\frac{\partial}{\partial x}\bigg{)}, \\
\widehat H_2 &=& -\frac{\hbar^2}{3x\sigma^{\frac{4}{3}}}\frac{\partial^2}{\partial \phi^2}, \\
\widehat H_3 &=&  \frac{2i\hbar } {3\sigma}\bigg{(}\frac{3\alpha_0}{\phi^2}-\frac{2{\gamma_0}\phi x^2}{\sigma^{\frac{4}{3}}} \bigg{)} \frac{\partial}{\partial \phi} +\frac{2i\hbar}{\sigma} \bigg{(}\frac{\gamma_0 x^2}{3 \sigma^\frac{4}{3}}-{\alpha_0\over\phi^3}\bigg{)},\\
\widehat V_e &=& V_e.
\end{eqnarray}
Now, let us consider the first term,

\be \label{1}\int \big{(}\widehat H_1\Psi\big{)}^*\Psi dx= -\frac{\hbar^2\phi}{54\beta_0}\int\bigg{(}\frac{1}{x}\frac{\partial^2\Psi^*}{\partial x^2} +\frac{n}{x^2}\frac{\partial\Psi^*}{\partial x}\bigg{)}\Psi dx=-\frac{\hbar^2\phi}{54\beta_0}\int\bigg{(}\frac{\Psi}{x}\frac{\partial^2\Psi^*}{\partial x^2} +\frac{n\Psi}{x^2}\frac{\partial\Psi^*}{\partial x}\bigg{)} dx.\ee
Twice we integrate the above integral \eqref{1} by parts, and drop the first term due to fall-of condition, to obtain,

    \be\label{H1} \int \big{(}\widehat H_1\Psi\big{)}^*\Psi dx=-\frac{\hbar^2\phi}{54\beta_0}\int \Psi^*\bigg{[}\frac{1}{x}\frac{\partial^2\Psi}{\partial x^2} -\frac{n+2}{x^2}\frac{\partial\Psi}{\partial x}+\frac{2(n+1)}{x^3}\Psi\bigg{]}dx. \ee
Under the particular choice of the operator ordering index viz., $n=-1$, we can express the above integral (\ref{H1}) as,

\be\label{H1.1} \int \big{(}\widehat H_1\Psi\big{)}^*\Psi dx=-\frac{\hbar^2\phi}{54\beta_0}\int \Psi^*\bigg{[}\frac{1}{x}\frac{\partial^2\Psi}{\partial x^2} -\frac{1}{x^2}\frac{\partial\Psi}{\partial x}\bigg{]}dx=\int \Psi^*\widehat H_1\Psi dx. \ee
rendering $\widehat H_1$ hermitian, of-course for a particular choice of operator ordering parameter $n=-1$. Further, $\widehat H_2$ is typically hermitian. We therefore take up the third term viz., $\widehat H_3,$

\be \int (\widehat H_3 \Psi)^*\Psi d\phi= -\frac{2i\hbar } {3\sigma}\int\bigg{(}\frac{3\alpha_0}{\phi^2}-\frac{2{\gamma_0}\phi x^2}{\sigma^{\frac{4}{3}}}\bigg{)} \frac{\partial\Psi^*}{\partial \phi}\Psi d\phi -\frac{2i\hbar}{\sigma}\int \bigg{(}\frac{\gamma_0 x^2}{3 \sigma^\frac{4}{3}}-{\alpha_0\over\phi^3}\bigg{)} \Psi^*\Psi d\phi.\ee
Under integration by parts and dropping the integrated out terms due to fall-of condition as usual, we obtain,

\be \int (\widehat H_3 \Psi)^*\Psi d\phi= \frac{2i\hbar } {3\sigma}\int\Psi^* \bigg{(}\frac{3\alpha_0}{\phi^2}-\frac{2{\gamma_0}\phi x^2}{\sigma^{\frac{4}{3}}}\bigg{)} \frac{\partial\Psi}{\partial \phi} d\phi +\frac{2i\hbar}{\sigma}\int \bigg{(}\frac{\gamma_0 x^2}{3 \sigma^\frac{4}{3}}-{\alpha_0\over\phi^3}\bigg{)} \Psi^*\Psi d\phi=\int \Psi^*\widehat H_3 \Psi d\phi,\ee
again rendering the fact that the effective Hamiltonian $\widehat H_3$ is hermitian too. Thus, $\widehat H_e$ finally turns out to be a hermitian operator. The hermiticity of $\widehat H_e$ now allows one to write the continuity equation, as,
\begin{center}
    \be \frac{\partial\rho}{\partial\sigma}+\nabla. \textbf{J}=0, \ee
\end{center}
which requires to find $\frac{\partial\rho}{\partial\sigma}$, where, $\rho=\Psi^*\Psi$.  A little algebra leads to the following equation:

\be\begin{split} \frac{\partial\rho}{\partial\sigma}=&-\frac{\partial}{\partial x}\bigg{[}\frac{i\hbar\phi}{54\beta_0 x}\big{(}\Psi\Psi^*_{,x}-\Psi^*\Psi_{,x} \big{)}\bigg{]}-\frac{\partial}{\partial\phi}\bigg{[}\frac{i\hbar}{3x\sigma^{\frac{4}{3}}} \big{(}\Psi \Psi^*_{,\phi}-\Psi^*\Psi_{,\phi} \big{)}-\frac{2}{3\sigma}\Big{(}\frac{3\alpha_0}{\phi^2}-\frac{2{\gamma_0}\phi x^2}{\sigma^{\frac{4}{3}}} \Big{)}\Psi^*\Psi\bigg{]} \\&-\frac{i\hbar\phi}{54\beta_0}\frac{(n+1)}{x^2}\big{(}\Psi\Psi^*_{,x}-\Psi^*\Psi_{,x} \big{)}, \end{split} \ee
where, we have used the symbols for derivatives as, $\frac{\partial\Psi}{\partial x} = \Psi_{,x}, \frac{\partial\Psi}{\partial \phi} = \Psi_{,\phi}$ etc. Clearly, the continuity equation can be written, only under the same earlier choice $n=-1$ as,

\be \frac{\partial\rho}{\partial\sigma}+\frac{\partial{J}_x}{\partial x}+\frac{\partial{J}_\phi}{\partial \phi}=0.
\ee
In the above, $\rho=\Psi^* \Psi$ and $\textbf{J}=({J}_x, {J}_{\phi}, 0)$ are the probability density and current density respectively,
where,

\begin{eqnarray}
{J}_x &=& \frac{i\hbar\phi}{54\beta_0 x}\big{(}\Psi\Psi^*_{,x}-\Psi^*\Psi_{,x} \big{)}, \\
{J}_{\phi} &=&  \frac{i\hbar}{3x\sigma^{\frac{4}{3}}} \big{(}\Psi \Psi^*_{,\phi}-\Psi^*\Psi_{,\phi} \big{)}-\frac{2}{3\sigma}\Big{(}\frac{3\alpha_0}{\phi^2}-\frac{2{\gamma_0}\phi x^2}{\sigma^{\frac{4}{3}}}\Big{)}\Psi^*\Psi.
\end{eqnarray}
Here, the variable $\sigma$ plays the role of internal time parameter, as already mentioned. Thus, the operator ordering index has been fixed to $n = -1$, to establish the physical requirement that the Hamiltonian operator has to be hermitian.

\subsection{Semiclassical approximation:}

Semiclassical approximation is essentially a method of finding an approximate wavefunction associated with a quantum equation. If the integrand in the exponent of the semiclassical wavefunction is imaginary, then the approximate wave function is oscillatory, and falls within the classical allowed region. Thus, a quantum theory is justified, when the semiclassical approximation is working, i.e. admits classical limit. In such case most of the important physics lies in the classical action. A quantum theory therefore, may only be accepted as viable, if it admits and also found to be well-behaved under, an appropriate semiclassical approximation. To justify the present quantum equation (\ref{qh2}) in the context of the acid test mentioned above, we therefore need to study its behaviour under certain appropriate semiclassical limit in the standard WKB approximation. For this purpose it is much easier to handle the equation (\ref{qh1}), and express it in the following form,

\be \label{qh1s} \begin{split}\bigg{[}-\frac{\hbar^2{\sqrt z\phi}}{36\beta_{0} x}\bigg{(}\frac{\partial^2}{\partial x^2} +\frac{n}{x} \frac{\partial}{\partial x}\bigg{)}-\frac{\hbar^2}{2xz^{\frac{3}{2}}}\frac{\partial^2}{\partial \phi^2}&-{i\hbar} \frac{\partial} {\partial z}+\frac{i\hbar } {z}\bigg{(}\frac{3\alpha_0}{\phi^2}-\frac{2\gamma_0x^2\phi}{z^2}\bigg{)}\frac{\partial}{\partial\phi}-\frac{3i\hbar}{z}\bigg{(}\frac{\alpha_0} {\phi^3}+\frac{\gamma_0x^2}{3z^2}\bigg{)} +\mathcal{V}\bigg{]}\Psi=0,
\end{split}\ee
where,

\be\begin{split} \mathcal{V}=\bigg[\frac{9\alpha_{0}^2x}{2\sqrt{z}\phi^4}-\frac{6\alpha_{0}\gamma_{0}x^3}{z^\frac{5}{2}\phi}+\frac{2\gamma_{0}^2 x^5 \phi^2}{z^\frac{9}{2}}+\frac{3\alpha_{0} x}{2\sqrt{z}\phi}+\frac{\lambda^2 z^\frac{3}{2} \phi^2}{2 x}+{\Lambda M^2_P z^{3\over 2}\over x}\bigg].\end{split} \ee
Equation (\ref{qh1s}) may be treated as time independent Schr{\"o}dinger equation with three variables ($x$, $z$, $\phi$), and therefore, we seek the solution of equation (\ref{qh1s}) as usual, in the following form \footnote{Although, in the semiclassical approximation, the amplitude should be treated as slowly varying function with respect to the phase, however, the Hamilton-Jacobi equation remains the same and the semiclassical wavefunction, at least up to the first order approximation that we perform, remains unaltered. Therefore, to avoid complication, we consider the amplitude to be a constant.},

\be\label{Psi}\Psi = \Psi_0e^{\frac{i}{\hbar}S(x,z,\phi)}\ee
and expand $S$ in power series of $\hbar$ as,

\be\label{S} S = S_0(x,z,\phi) + \hbar S_1(x,z,\phi) + \hbar^2S_2(x,z,\phi) + .... .\ee
Now inserting the expressions (\ref{Psi}) and (\ref{S}) together with appropriate derivatives ($\Psi_{,x}, \Psi_{,xx}, \Psi_{,\phi}, \Psi_{,\phi\phi}, \Psi_{,z}$ etc.) in equation (\ref{qh1s}) and equating the coefficients of different powers of $\hbar$ to zero, one obtains the following set of equations (upto second order),

\be\begin{split}
&\frac{\sqrt z\phi}{36\beta_0 x}S_{0,x}^2 + \frac{S_{0,\phi}^2}{2xz^{\frac{3}{2}}} + S_{0,z}-\frac{1} {z}\bigg{(}\frac{3\alpha_0}{\phi^2}-\frac{2\gamma_0x^2\phi}{z^2}\bigg{)} S_{0,\phi} + \mathcal{V}(x,z,\phi) = 0, \label{S0}\end{split}\ee
\be\begin{split}
-\frac{i\sqrt z\phi}{36\beta_0 x}S_{0,xx} - \frac{in\sqrt z\phi}{36\beta_0 x^2}S_{0,x} - \frac{i S_{0,\phi\phi}}{2xz^{\frac{3}{2}}} + S_{1,z} &+ \frac{ \sqrt z \phi S_{0,x}S_{1,x}}{18\beta_0  x}
 + \frac{S_{0,\phi}S_{1,\phi}}{xz^{\frac{3}{2}}}-\frac{3i}{z}\bigg{(}\frac{\alpha_0}{\phi^3}+\frac{\gamma_0 x^2}{3 z^2}\bigg{)}\\&-\frac{1} {z}\bigg{(}\frac{3\alpha_0}{\phi^2}-\frac{2\gamma_0x^2\phi}{z^2}\bigg{)} S_{1,\phi} = 0, \label{S1}\end{split}\ee
\be\begin{split} -\frac{i \sqrt z \phi S_{1,xx}}{36\beta_0 x} - \frac{i n\sqrt z \phi S_{1,x}}{36\beta_0 x^2}+\frac{\sqrt z\phi}{36\beta_0 x}\Big({S_{1,x}^2+2 S_{0,x}S_{2,x}}\Big)& + \frac{1}{{2xz^{\frac{3}{2}}}}\Big({ S_{1,\phi}^2+2S_{0,\phi}S_{2,\phi}}\Big) -
\frac{i S_{1,\phi\phi}}{2xz^{\frac{3}{2}}}\\& + S_{2,z}-\frac{1} {z}\bigg{(}\frac{3\alpha_0}{\phi^2}-\frac{2\gamma_0x^2\phi}{z^2} \bigg{)} S_{2,\phi}= 0,\label{S2}
\end{split}\ee
which are to be solved successively to find $S_0(x,z,\phi),\; S_1(x,z,\phi)$ and $S_2(x,z,\phi)$ and so on. Now identifying $S_{0,x}$ as $p_x$, $S_{0,z}$ as $p_z$ and $S_{0,\phi}$ as $p_{\phi}$ one can recover the classical Hamiltonian constraint equation $H_c = 0$, presented in equation (\ref{Hc}) from equation (\ref{S0}). Thus, $S_{0}(x, z)$ can now be expressed as,

\be\label{S_0} S_0 = \int p_z dz + \int p_x dx + \int p_\phi d\phi, \ee
apart from a constant of integration which may be absorbed in $\Psi_0$. The integrals in the above expression can be evaluated using the classical solution for $k = 0$ presented in equation \eqref{aphi}, (\ref{param}), \eqref{betagamma}, and the definitions of $p_z$ given in (\ref{pz}), $p_{\phi}$ in (\ref{pph})) and also using the relation $p_x = {Q}$. For the last expression for the momentum, it is required to recall the expression for ${Q}$ given in (\ref{pQ}), remembering the relation, $x = \dot z$, where, $z = a^2$. Further, we choose $n = -1$, since probability interpretation holds only for such choice of $n$. Now, using classical de-Sitter solution together with the expressions for $\alpha({\phi})$ given in \eqref{param}, and that of $\beta(\phi)$, $\gamma(\phi)$ presented in \eqref{betagamma}: the variable $x (= \dot z)$ along with all the expressions of momenta, viz. $p_x$, $p_z$ and $p_\phi$ may be expressed in term of $x$, $z$ and $\phi$ as,

\begin{subequations}\begin{align}
&\label{allp} \hspace{5.0 cm}\alpha'=-\frac{\alpha_0}{\phi^2},\\
&\hspace{5.0 cm}x = 2{\lambda} z, \\
&\hspace{5.0 cm}p_x = \frac{36\beta_0\lambda x}{a_0\phi_0},\\
&\hspace{5.0 cm}p_z = -\frac{9\alpha_0\lambda z}{a_0\phi_0}-\frac{144\beta_0\lambda^3 z}{a_0\phi_0}+\frac{24\gamma_{0}a_{0}^2\phi_{0}^2{\lambda^3}}{\sqrt{z}},\\
&\hspace{5.0 cm}p_\phi = \frac{6\alpha_0 a_0^3\phi_0^3{\lambda}}{\phi^5}-\frac{16\gamma_0a_0^3\phi_0^3\lambda^3}{\phi^2}-\frac{a_0^3\phi_0^3\lambda}{\phi^2},
                                \end{align}\end{subequations}
such that, the integrals in (\ref{S_0}) are evaluated as,
\begin{subequations}\begin{align}
&\label{px2pphi}\hspace{5.0 cm}\int p_x dx =\frac{18\beta_0\lambda x^2}{a_0\phi_0}; \\
&\hspace{5.0 cm}\int p_z dz =-\frac{9\alpha_0\lambda z^2}{2a_0\phi_0}-\frac{72\beta_0 \lambda^3 z^2}{a_0\phi_0}+48\gamma_{0}a_{0}^2\phi_{0}^2\lambda^3 \sqrt{z}; \\
&\hspace{5.0 cm}\int p_\phi d\phi = -\frac{3\alpha_0 a_0^3\phi_0^3{\lambda}}{2\phi^4}+\frac{16\gamma_0a_0^3\phi_0^3\lambda^3}{\phi}+\frac{a_0^3\phi_0^3\lambda}{\phi}.
\end{align}\end{subequations}
\noindent
Hence, explicit form of $S_0$ in terms of $z$ is found as,

\be\label{S0f}\hspace{5.0 cm}S_0 = -\frac{6\alpha_0\lambda z^2}{a_0\phi_0}+16\gamma_{0}a_{0}^2\phi_{0}^2\lambda^3\sqrt{z}.\ee
For consistency, one can trivially check that the expression for $S_0$ (\ref{S0f}) so obtained, satisfies equation (\ref{S0}) identically. In fact it should, because, equation (\ref{S0}) coincides with Hamiltonian constraint equation (\ref{Hc}) for $k = 0$. This unambiguously establishes the fact that equation (\ref{S0}) is the Hamilton-Jacobi equation, while $S_0$ is the Hamilton-Jacobi function. Moreover, one can also compute the zeroth order on-shell action (\ref{Sc3}). Using the classical solution (\ref{aphi}), \eqref{param} and \eqref{betagamma}, one may express all the variables in terms of $t$ and substitute in the action (\ref{Sc3}) to obtain,

\be\label{Acl}A=A_{cl}=\int\left[-\frac{24\alpha_0 a_0^3\lambda^2}{\phi_0}e^{{4\lambda}t}+16\gamma_0a_0^3\phi_0^2{\lambda}^4e^{\lambda t}\right]dt. \ee
Integrating we have,

\be\label{OnSh} A=A_{cl} = -\frac{6\alpha_0a_0^3\lambda}{\phi_0}e^{{4\lambda}t}+16\gamma_0a_0^3\phi_0^2{\lambda}^3e^{\lambda t},\ee
which is the same as the Hamilton-Jacobi function obtained in (\ref{S0f}), since $z = a_0^2 e^{2\lambda t}$. This proves consistency of the present approach towards semiclassical approximation. At this end, the wave function reads as,

\be\label{psif} \Psi = \Psi_0 e^{\frac{i}{\hbar}\left[ -\frac{6\alpha_0\lambda z^2}{a_0\phi_0}+16\gamma_{0}a_{0}^2\phi_{0}^2\lambda^3\sqrt{z}\right]},\ee\\
exhibiting oscillatory behaviour.\\

\noindent
\textbf{First order approximation:}\\

\noindent
Under the choice, $n=-1$, equation (\ref{S1}) can be expressed as,

\be \begin{split}
-\frac{\sqrt z\phi}{36\beta_0 x}\bigg({}i S_{0,xx}-2S_{0,x}S_{1,x} - \frac{i}{ x}S_{0,x}\bigg{)} &- \frac{1}{2xz^{\frac{3}{2}}} \bigg{(}i{S_{0,\phi\phi}}-2S_{0,\phi}S_{1,\phi}\bigg{)}-\frac{1}{z}\bigg{(}\frac{3\alpha_0}{\phi^2}-\frac{2\gamma_0x^2\phi}{z^2}\bigg{)} S_{1,\phi}\\& -{3i\over z}\bigg{(}{\alpha_0\over\phi^3}+\frac{\gamma_0x^2}{3z^2}\bigg{)}+ S_{1,z} = 0. \label{S11}\end{split}\ee
One can now compute appropriate derivatives of the expression of $S_0$ given in (\ref{S0f}), to obtain the expression of $S_{1,z}$ from the above equation \eqref{S11} as,

\be\label{S1z} S_{1,z} = i \left(\frac{{C_1\sqrt z} +\frac{C_2}{z}+\frac{C_3}{z^\frac{5}{2}}}{D_1z^\frac{3}{2}+\frac{D_2}{z^\frac{3}{2}}+D_3}\right),\ee
where, $C_1=-\frac{27\alpha_0}{ a_0^3\phi_0^3},~~C_2=12\gamma_0 \lambda^2,~~ C_3=-\frac{\gamma_0 a_0^3 \phi_0^3}{24 \beta_0},~~D_1=-{18\alpha_0\over a_0^3 \phi_0^3},~~D_2=\frac{\gamma_0 a_0^3 \phi_0^3}{18\beta_0}$, and $D_3=\left(1-\frac{\alpha_0}{12\beta_0\lambda^2}\right)$ are all constants. The above equation \eqref{S1z} may be integrated in principle, and  $S_1$ may be expressed in the form,

\be S_1 = i F(z).
\ee
Therefore the wave function up to first-order approximation reads as,

\be \Psi = \Psi_{01} e^{\frac{i}{\hbar}\left[-\frac{6\alpha_0\lambda z^2}{a_0\phi_0}+16\gamma_{0}a_{0}^2\phi_{0}^2\lambda^3\sqrt{z} \right]},\ee
where,
\be \Psi_{01}=\Psi_0e^{-F(z)},\ee
which only tells upon the pre-factor keeping the exponent part unaltered. We have therefore exhibited a technique to find the semiclassical wavefunction, on-shell. One can proceed further to find higher order approximations. Nevertheless, it is clear that higher order approximations too, in no way would affect the form (exponent) of the semiclassical wavefunction, which has been found to be oscillatory around the classical inflationary solution. Since, the semiclassical wavefunction exhibits oscillatory behaviour around classical de-Sitter solution \eqref{aphi}, it is therefore strongly peaked around classical inflationary solutions. Thus we prove that the quantum counterpart of the action \eqref{A} produces a reasonably viable theory. It is important to mention that the semiclassical approximation is validated only if it occurs at sub-Planckian energy scale. Indeed it is so, as we shall find in the next section that the energy scale is $\mathrm{H}_* \approx 10^{-5} M_P$, where $\mathrm{H}_*$ is the Hubble parameter which determines the energy scale during inflationary regime.

\section{\bf{Inflation under Slow Roll Approximation:}}

Having proved the viability of the action \eqref{A} in the quantum domain, we now proceed to test inflation with currently released data sets in this regard \cite{pd1, pd2}. Inflation is a quantum phenomena, which was initiated sometime between ($10^{-36}$ and $10^{-26}$) sec., after gravitational sector transits to the classical domain. To be more specific, inflation is a quantum theory of perturbations on top of a classical background, which means the energy scale of the background must be much below Planck scales. There are also recent hints from the string theory swampland that the energy scale must be rather low for inflation. In the previous sub-section we have mentioned that if a quantum theory admits a viable semiclassical approximation, then most of the important physics may be extracted from the classical action itself. Clearly, the above  semiclassical approximation is validated if the energy scale of inflation is below Planck's scale. For a complicated theory such as the present one, the computation of inflationary parameters is of-course a very difficult job. However, we follow a unique technique to make things look rather simple. Let us first rearrange the ($^0_0$) and the $\phi$ variation equations of Einstein, viz., (\ref{00}) and (\ref{phivariation}) respectively as,

\be\label{00H}\begin{split}  6\alpha \mathrm{H}^2 + 6\alpha'{\dot\phi}\mathrm{H} + 36\beta \mathrm{H}^4 \bigg{[}4\bigg{(}1+\frac{\dot {\mathrm{H}}}{\mathrm{H}^2}\bigg{)}&+4\frac{\dot{ \mathrm{H}}}{\mathrm{H}^2}\bigg{(}1+\frac{\dot{\mathrm{H}}}{\mathrm{H}^2}\bigg{)} +2\bigg{(}\frac{\ddot {\mathrm{H}}}{\mathrm{H}^3}-2\frac{{\dot {\mathrm{H}}}^2}{\mathrm{H}^4}\bigg{)}-\bigg{(}1+\frac{\dot {\mathrm{H}}} {\mathrm{H}^2} \bigg{)}^2-3\bigg{]}\\& + 72\beta'{\dot\phi} \mathrm{H}^3\bigg{[}\bigg{(}1+\frac{\dot {\mathrm{H}}}{\mathrm{H} ^2}\bigg{)}+ 1\bigg{]} + 24\gamma'\dot\phi \mathrm{H}^3 = V+\Lambda M^2_P+\frac{\dot\phi^2}{2}, \end{split} \ee

\be \label{phivar} \begin{split}& \ddot\phi +3\mathrm{H}\dot\phi + V' = 6\alpha'\mathrm{H}^2\bigg{[}\bigg{(}1+\frac{\dot {\mathrm{H}}} {\mathrm{H}^2} \bigg{)}+1\bigg{]}+36\beta' \mathrm{H}^4\bigg{[} \bigg{(}1+\frac{\dot{\mathrm{H}}}{\mathrm{H}^2}\bigg{)}^2+2\bigg{(}1+\frac{\dot {\mathrm{H}}}{\mathrm{H}^2}\bigg{)}+1\bigg{]}+24\gamma'\mathrm{H}^4\bigg{(}1+\frac{\dot {\mathrm{H}}}{\mathrm{H}^2}\bigg{)},
\end{split}\ee
where, $\mathrm{H}={\dot a\over a}$ denotes the expansion rate, which is assumed to be slowly varying. Note that we have already presented inflationary solutions in \eqref{aphi} of the classical field equations (\ref{00}) and (\ref{phivariation}) in standard de-Sitter form. Now, instead of standard slow roll parameters, we introduce a hierarchy of Hubble flow parameters \cite{SR, CH1, CH2, CH3, CH4, jhep, We1, We2} in the following manner, which appears to be much suitable and elegant to handle higher order theories. Firstly, the background evolution of the theory under consideration is described by a set of horizon flow functions (the behaviour of Hubble distance during inflation) starting from,

\be \label{dh}\epsilon_0=\frac{d_{\mathrm{H}}}{d_{\mathrm{H}_i}},
\ee
where, $d_{\mathrm{H}} = \mathrm{H}^{-1}$ is the Hubble distance, also called the horizon in our chosen units. We use suffix $i$ to denote the era at which inflation was initiated. Now hierarchy of functions is defined in a systematic way as,

\begin{center}
    \be \label{el} \epsilon_{l+1}=\frac{d\ln|\epsilon_l|}{d \mathrm{N}},~~l\geq 0. \ee
\end{center}
In view of the definition $\mathrm{N}=\ln{\frac{a}{a_i}}$, implying $\mathrm{\dot N}=\mathrm{H}$, one can compute $\epsilon_1=\frac{d\ln{d_{\mathrm{H}}}}{d\mathrm{N}},$ which is the logarithmic change of Hubble distance per e-fold expansion $\mathrm{N}$, and is the first slow-roll parameter: $\epsilon_1=\dot{d_{\mathrm{H}}}=-\frac{\dot {\mathrm{H} }}{\mathrm{H}^2}$. This ensures that the Hubble parameter almost remains constant during inflation $\epsilon_1\ll 1$. The above hierarchy allows one to compute $\epsilon_2=\frac{d\ln{\epsilon_1}}{d\mathrm{N}}=\frac{1}{\mathrm{H}}\frac{\dot\epsilon_1}{\epsilon_1},$ which implies $\epsilon_1\epsilon_2=d_{\mathrm{H}} \ddot{d_{\mathrm{H}}} =-\frac{1}{\mathrm{H}^2}\left(\frac{\ddot {\mathrm{H}}}{\mathrm{H}}-2\frac{\dot {\mathrm{H}}^2}{\mathrm{H}^2}\right)$. In the same manner higher slow-roll parameters may be computed. Equation (\ref{el}) essentially defines a flow in space with cosmic time being the evolution parameter, which is described by the equation of motion

\be\label{el1}\epsilon_0\dot\epsilon_l-\frac{1}{d_{\mathrm{H}_i}}\epsilon_l\epsilon_{l+1}=0,~~~~l\geq 0.\ee
In view of the slow-roll parameters, equations (\ref{00H}) and (\ref{phivar}) may therefore be expressed as,

\be \label{hir1}\begin{split} -6\alpha \mathrm{H}^2-6\alpha'\dot\phi\mathrm{H}-36\beta \mathrm{H}^4\left[3\big(1-\epsilon_1\big)^2-2\big(1+\epsilon_1 \epsilon_2 \big)-1\right]&-72\beta'\dot\phi\mathrm{H} ^3\left[\big(1- \epsilon_1\big)+1\right]\\& -24\gamma'\dot\phi \mathrm{H}^3+\Big(\frac{\dot\phi^2}{2}+V+\Lambda M^2_P\Big)=0, \end{split}\ee
and
\be \label{phi2} \ddot\phi +3\mathrm{H}\dot\phi=-V'+6\alpha'\mathrm{H}^2\left[3-\big(1+\epsilon_1\big)\right]+{36\beta'\mathrm{H}^4}\left[\big(1-\epsilon_1\big)^2+2\big(1-\epsilon_1\big)+1\right]+24\gamma' \mathrm{H}^4\big(1- \epsilon_1\big), \ee
respectively, which may therefore be approximated using the slow roll hierarchy to,

\be \label{hir11}\begin{split} 6\alpha \mathrm{H}^2 = \frac{\dot\phi^2}{2} + \left[V+\Lambda M^2_P - \big(6\alpha'\dot\phi\mathrm{H} + 144\beta'\dot\phi\mathrm{H} ^3 + 24\gamma'\dot\phi \mathrm{H}^3\big)\right], \end{split}\ee
and
\be \label{phi22} \ddot\phi +3\mathrm{H}\dot\phi + \left[V' - \big(12\alpha'\mathrm{H}^2+{144\beta'\mathrm{H}^4}+24\gamma' \mathrm{H}^4\big)\right] = 0. \ee
Before imposing the standard slow roll conditions, viz. $|\ddot \phi| \ll 3\mathrm{H}|\dot \phi|$ and $\dot\phi^2 \ll V(\phi)$, we try to reduce equations \eqref{hir11} and \eqref{phi22} in a much simpler form. For example redefining the potential as,

\be\label{U1} U = V -12\mathrm{H}^2(\alpha+12\mathrm{H}^2\beta+2\mathrm{H}^2\gamma),\ee
equation (\ref{phi22}) takes the standard form of Klien-Gordon Equation,

\be \label{KG}\begin{split} \ddot\phi +3\mathrm{H}\dot\phi + U'=0 ;\end{split}. \ee
In view of the reduced equation \eqref{KG}, it is now quite apparent that the evolution of the scalar field is driven by the re-defined potential gradient $U' = {dU\over d\phi}$, subject to the damping by the Hubble expansion $3 \mathrm{H} \dot \phi$, as in the case of single field equation, while the potential $U(\phi)$ carries all the information in connection with the coupling parameters of generalised higher order action under consideration. Further, assuming

\be\label{U2} U = V + M^2_P \Lambda  - 6\mathrm{H}\dot \phi\left(\alpha' + 24 \mathrm{H}^2 \beta' + 4 \mathrm{H}^2 \gamma'\right),\ee
equation \eqref{hir11} may be reduced to the following simplified form, viz,

\be\label{Fried} 6\alpha\mathrm{H}^2 = \frac{\dot\phi^2}{2} + U(\phi).\ee
It is important to mention that, the two choices on the redefined potential $U(\phi)$ made in \eqref{U1} and \eqref{U2}, do not confront in any case and may be proved to be consistent as demonstrated underneath. During slow roll, the Hubble parameter $\mathrm{H}$ almost remains unaltered. Thus replacing $\mathrm{H}$ by $\lambda$, and using the forms of the parameters $\alpha(\phi)$ presented in \eqref{param}, along with $\beta(\phi)$ and $\gamma(\phi)$ assumed in \eqref{betagamma}, the two relations \eqref{U1} and \eqref{U2} lead to the following first order differential equation on $\phi$,

\be \label{Consistency} \left(\frac{\phi^3 - 6\alpha_0}{\lambda^2\phi^4-2\Lambda M^2_P\phi^2}\right) d\phi = \frac{1}{2\lambda} dt,\ee
which can immediately be solved to yield,

\be \label{phidecay}\begin{split}&\phi (t)=\text{InverseFunction}\left[\frac{\log \left(2 \Lambda -\text{$\#$1}^2 \lambda ^2\right)}{2 \lambda ^2}+\frac{3 \alpha _0 \lambda  \tanh ^{-1}\left(\frac{\text{$\#$1} \lambda }{\sqrt{2} \sqrt{\Lambda }}\right)}{\sqrt{2} \Lambda ^{3/2}}-\frac{3 \alpha _0}{\text{$\#$1} \Lambda }\&\right]\left[c_1+\frac{t}{2 \lambda }\right],\end{split}\ee
where, $c_1$ is a constant of integration. Obviously, the solution is complicated to explore the behaviour of $\phi$ against $t$. However, we shall show later, separately both in the cases without and with the cosmological constant (under certain reasonable assumption based on the data set), that indeed $\phi$ falls-of with time, as required.\\

Having proven consistency of our assumptions, we enforce the standard slow-roll conditions $\dot\phi^2\ll U$ and $|\ddot\phi|\ll 3\mathrm{H}|\dot\phi|$, on equations (\ref{Fried}) and (\ref{KG}), which thus finally reduce to,

\be\label{H2}{6\alpha}\mathrm{H}^2\simeq U, \ee
and
\be\label{Hphi} 3\mathrm{H}\dot\phi \simeq - U'. \ee
Now, combining equations (\ref{H2}) and (\ref{Hphi}), it is possible to show that the potential slow roll parameter $\epsilon$ equals the Hubble slow roll ($\epsilon_1$) parameter under the condition,

\be\label{SR} \epsilon = - {\dot {\mathrm{H}}\over \mathrm{H}^2} = \alpha\left({U'\over U}\right)^2 - \alpha'\left({U'\over U}\right);\hspace{0.4 cm} \eta = 2 \alpha \left({U''\over U}\right),\ee
while $\eta$ remains unaltered. Further, since $\frac{\mathrm{H}}{\dot\phi}=-{U\over2\alpha U'}$, therefore, the number of e-folds at which the present Hubble scale equals the Hubble scale during inflation, may be computed as usual in view of the following relation:

\be\label{Nphi} \mathrm{N}(\phi)\simeq \int_{t_i}^{t_f}\mathrm{H}dt=\int_{\phi_i}^{\phi_f}\frac{\mathrm{H}}{\dot\phi}d\phi\simeq \int_{\phi_f}^{\phi_i}\Big{(}\frac{U}{2\alpha U'}\Big{)}d\phi,\ee
where, $\phi_i$ and $\phi_f$ denote the values of the scalar field at the beginning $(t_i)$ and the end $(t_f)$ of inflation. Thus, slow roll parameters reflect all the interactions, as exhibited earlier \cite{GB1, GB2, GB3}, but here only via the redefined potential $U(\phi)$.\\

\subsection{\bf{Inflation without cosmological constant:}}

Let us first consider cosmological constant $\Lambda=0$, so that the potential $V(\phi) = {1\over 2}\lambda^2\phi^2$, while the forms of $\alpha(\phi)$ \eqref{param}, $\beta(\phi)$ and $\gamma(\phi)$ \eqref{betagamma} satisfy classical de-Sitter solutions as well. Now, in order to compute inflationary parameters numerically, it is necessary to find the form of the re-defined potential $U(\phi)$. As mentioned, during inflation the Hubble parameter remains almost constant, and therefore while computing $U(\phi)$, one can replace it by the constant $\lambda$, without any loss of generality. Thus,

\be 12\mathrm{H}^2\left(\alpha+12\mathrm{H}^2\beta+2\mathrm{H}^2\gamma\right) \approx -\frac{\mathrm{H}^2\phi^2}{2},~~\mathrm{such~ that},~~ U = {1\over 2} {m^2\phi^2},~~\mathrm{where},~~ m^2 = \lambda^2+\mathrm{H}^2 \approx 2\lambda^2.\ee
\noindent
In view of the above form of $U$, we find the following relation from \eqref{Fried},

\be 6{\alpha_0\over\phi}\mathrm{H}^2={1\over 2}m^2\phi^2.\ee
Now choosing, $\alpha_0 =0.0078 M^3_P ,~ m^2 = 2\times10^{-12}M^2_P,~\phi_i=2.0 M_P$, it is possible to calculate from the above equation,
$\mathrm{H}^2 \approx 1.71\times 10^{-10}M^2_P$, and hence the energy scale of inflation ($\mathrm{H}_* \approx 10^{-5} M_P$) is sub-Planckian. Note that the inflationary parameters do not depend on $m^2$. Further, in view of the above quadratic form of the re-defined potential, the slow roll
parameters ($\epsilon,~\eta$) \eqref{SR} and the number of e-folding $\mathrm{N}$ \eqref{Nphi} take the following forms,

\be\label{epseta}\epsilon = \frac{6\alpha_0}{\phi^3},\hspace{0.5 cm}\eta = \frac{4\alpha_0}{\phi^3}, \hspace{0.5 cm}\mathrm{N} = {1\over 4 \alpha_0}\int_{\phi_f}^{\phi_i} \phi^2 d\phi = {1\over 12\alpha_0}(\phi_i^3 - \phi_f^3).\ee
Now, comparing expression for the primordial curvature perturbation on super-Hubble scales produced by single-field inflation $P_\zeta(k)$ with the primordial gravitational wave power spectrum $P_t(k)$, one obtains the tensor-to-scalar ratio for single-field slow-roll inflation $r = {P_t(k)\over P_\zeta(k)} = 16\epsilon$, while, the scalar tilt, conventionally defined as $n_s-1$, may be expressed as, $n_s = 1 - 6\epsilon + 2\eta$. In view of all these expressions we compute the inflationary parameters and present them for different values of the parameter $\alpha_0$  in table 1. We also present respective $n_s$ versus $r$ plots in figure 1.\\

\begin{figure}

   \begin{minipage}[h]{0.47\textwidth}
      \centering
      \begin{tabular}{|c|c|c|c|c|}
     \hline\hline
      $\alpha_0$ in $M_P^3$ & $\phi_f$ in $M_P$ & $n_s$ & $r$ & $\mathrm{N}$\\
      \hline
       0.0084 & 0.3694 & 0.9706 & 0.1008 & 79\\
      0.0082 & 0.3664 & 0.9713 & 0.0984 & 81\\
      0.0080& 0.3634 & 0.9720 & 0.0960 & 83\\
      0.0078 & 0.3604 & 0.9727 & 0.0936 &85\\
      0.0076 & 0.3573 & 0.9734 & 0.0912 & 87\\
      0.0074& 0.3541 & 0.9741 & 0.0888 & 90\\
      0.0072& 0.3509 & 0.9748 & 0.0864 & 92\\
      0.0070& 0.3476 & 0.9755 & 0.0840 & 95\\

      \hline\hline
    \end{tabular}
      \captionof{table}{Data set for the inflationary parameters taking $\phi_i=2.0 M_P$ and ~ varying $\alpha_0$.}
      \label{tab:table1}
   \end{minipage}%
   \hfill%
   \begin{minipage}[h]{0.47\textwidth}
      \centering
      \includegraphics[width=0.9\textwidth]{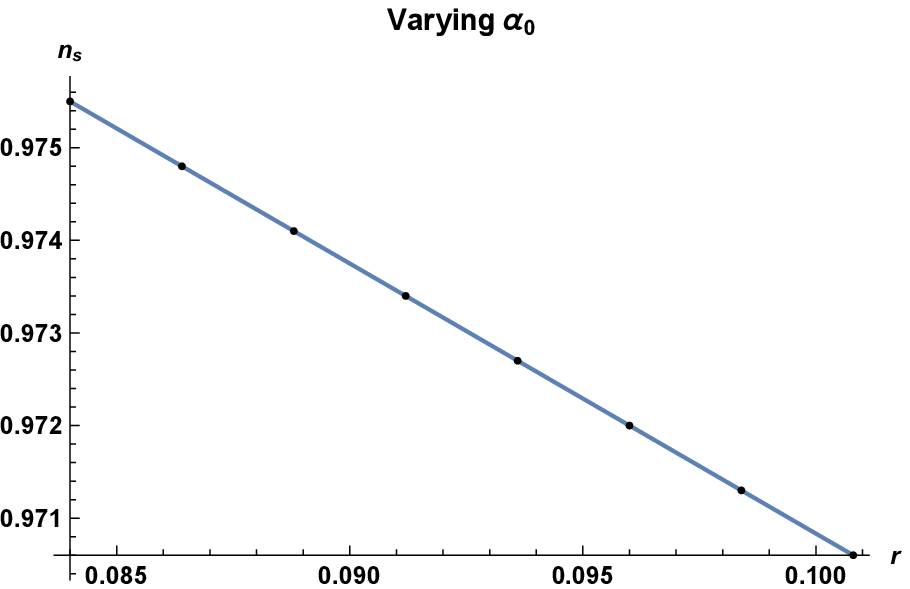}
      \caption{This plot depicts the variation of $n_s$ with $r$, varying $\alpha_0$.}
      \label{fig:fig1}
   \end{minipage}%

\end{figure}

\noindent
Table 1 depicts that under the variation of $\alpha_0$ within the range $0.0070 M_P^3\le\alpha_0 \le .0084M_P^3$, the spectral index of scalar perturbation and the scalar to tensor ratio lie within the range $0.970 \le n_s \le 0.976$ and $0.084\le r \le 0.101$ respectively, which show reasonably good agreement with the recently released data \cite{pd1, pd2}. For the sake of visualization we present the spectral index of scalar perturbation versus the scalar to tensor ratio plot, in figure 1.\\

\noindent
Although the theory under consideration is highly complicated, we have been able to reduce the system of field equations considerably to study inflation. In fact it is also possible to demonstrate that it does not suffer from graceful exit problem. In the absence of $\Lambda$ the equation \eqref{Consistency} reduces to,

\be \label{Con2} \left(\frac{\phi^3 - 6\alpha_0}{\lambda^2\phi^4}\right) d\phi = \frac{1}{2\lambda} dt,\ee
which can immediately be solved to yield

\be \label{phidecay1}\ln{\phi} + \frac{2\alpha_0}{\phi^3} = \frac{\lambda}{2}(t - t_0).\ee
Clearly, the scalar decays as $\phi \sim t^{-3}$ \eqref{phidecay1}, as depicted in figure 2, if $\phi~ (\le 2 M_P)$, is not too large, and quickly falls below Planck's mass, $\phi < M_p$.
 \begin{figure}
   \begin{minipage}[h]{0.47\textwidth}
      \centering
      \includegraphics[width=0.9\textwidth]{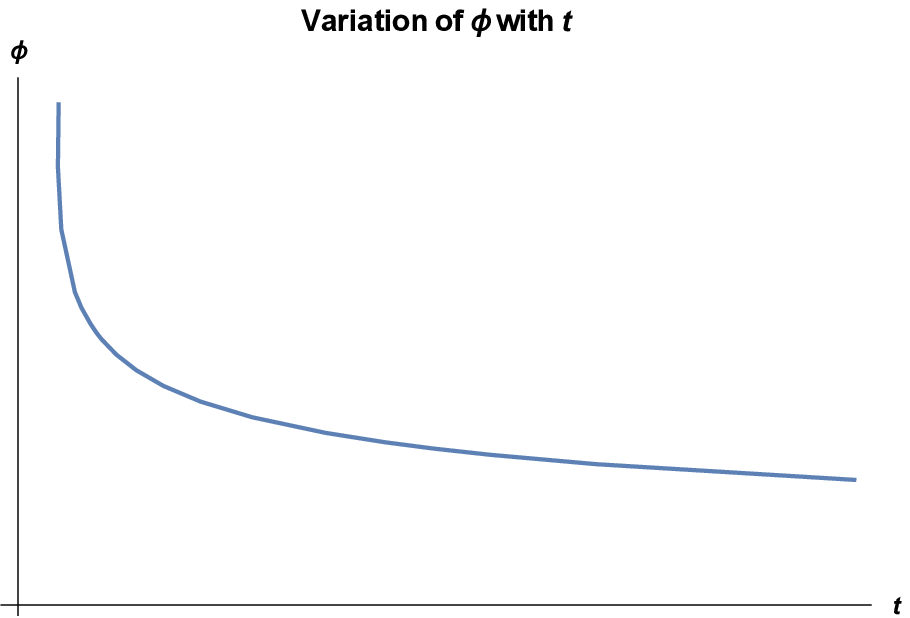}
      \caption{This plot depicts fall of $\phi$ with $t$.}
      \label{fig:fig2}
   \end{minipage}%

\end{figure}

\noindent
To exhibit the fact that as $\phi \ll M_p$ the field approaches an oscillatory solution, let us express equation \eqref{Fried} as,

\be 3\mathrm{H}^2 = {1\over 2\alpha}\left({1\over 2}\dot\phi^2 + {1\over 2} m^2 \phi^2\right),\ee
where, $U = {1\over 2} m^2 \phi^2$, and $m^2 \approx 2 \lambda^2$. In view of the expression of $\alpha(\phi) = {\alpha_0\over \phi}$, the above equation reads as,

\be \label{Hubble1} {3\mathrm{H}^2\over m^2} = {3\mathrm{H}^2\over 2\lambda^2} = {\phi\over 4\alpha_0}\left({\dot\phi^2\over 2\lambda^2} + \phi^2\right).\ee
Note that for single scalar field, the above equation leads to: $3\mathrm{H}^2 = {1\over 2M_p^2}(\dot\phi^2 + m^2 \phi^2)$. At the end of inflation, ${\phi\over 4\alpha_0} \sim {64\over M_p^2}$, according to the present data set. Once the Hubble rate falls below $\sqrt {2}\lambda$, this equation \eqref{Hubble1} may be approximated to,

\be \dot\phi^2 \approx -2\lambda^2 \phi^2,\ee
which exhibits oscillatory behaviour of $\phi \sim e^{i {\sqrt{2}\lambda}~ t}$. The field therefore starts oscillating many times over a Hubble time, driving a matter-dominated era at the end of inflation. Despite all these excellent features, the model runs into problem, since the number of e-folds varies within the range $79 \le \mathrm{N} \le 95$, being much larger than usual $\mathrm{N} \approx 60$. In fact, to fit the currently released datasets, $r \le .07$, the number of e-folds shoots up to $\mathrm{N} \approx 100$, which is too large. As a result, the universe cools down much more than the usual slow roll inflation, ending up with a very cold universe at the end of inflation, which could make filling the universe with matter rather difficult. This problem is somewhat related to dilaton stabilization issue. Note that in the present model, the same dilatonic field is responsible for inflation. It is known that generic string models suffer from dilaton runaway problem, since depending on the initial conditions, the field can acquire a large amount of kinetic energy and can easily overshoot any local minima in the potential. Such runaway behavior of the dilaton and other moduli fields prevents a viable inflation \cite{45}. Although the moduli stabilization issue has been studied by several authors earlier, so that it can successfully play the role of inflaton field without creating problems like reheating \cite{46,47,48}, nevertheless, since we have encountered problem, therefore let us consider next the role of cosmological constant in this respect.

\subsection{\bf{Inflation  with a cosmological constant:}}

It has been shown that an additive  cosmological constant$(M_P^2 \Lambda)$ does not affect the solutions to the classical field equations other than the fact that it only adds to the potential function. As mentioned, during inflation the Hubble parameter remains almost constant, and therefore while computing $U(\phi)$, one can replace it by the constant $\lambda$, without any loss of generality. Thus, taking into account the case 1a, we readily obtain,

\be\label{Mass} 12\mathrm{H}^2\left(\alpha+12\mathrm{H}^2\beta+2\mathrm{H}^2\gamma\right) \approx -\frac{\mathrm{H}^2\phi^2}{2},~~\mathrm{such~ that},~~ U = {1\over 2} {m^2\phi^2}-\Lambda M^2_P ,~~\mathrm{where},~~ m^2 = \lambda^2+\mathrm{H}^2 \approx 2\lambda^2.\ee
However, inflationary parameters $\epsilon,~\eta$ and the number of e-foldings $\mathrm{N}$ depend on the potential function itself and are likely to modify the situation. For example, in view of the above form of $U$, we obtain the following expression from equation \eqref{Fried},

\be \label{HM}6{\alpha_0\over\phi}\mathrm{H}^2={1\over 2}m^2\phi^2-\Lambda M^2_P.\ee
Now, for the above form of re-defined potential, the inflationary parameters read as,

\be\label{epseta2}\epsilon = \frac{4m^4\alpha_0\phi}{(m^2\phi^2-2\Lambda M^2_P)^2}+\frac{2m^2\alpha_0}{(m^2\phi^3-2\phi\Lambda M^2_P)},\hspace{0.6 cm}\eta = \frac{4m^2\alpha_0}{m^2\phi^3-2\phi\Lambda M^2_P}.\ee
\be\label{N}\mathrm{N} = {1\over 4\alpha_0}\int_{\phi_f}^{\phi_i}{(m^2\phi^2-2\Lambda M^2_P)\over m^2}d\phi = {1\over 12\alpha_0}(\phi_i^3 - \phi_f^3)-{\Lambda M^2_P\over 2m^2\alpha_0}(\phi_i-\phi_f).\ee
\noindent
We present data set in Table 2, which depicts that under the variation of $\alpha_0$ within the range $0.00012 M_P^3\le\alpha_0 \le 0.00019 M_P^3$, the spectral index of scalar perturbation and the scalar to tensor ratio lie within the range $0.967 \le n_s \le 0.979$ and $0.056\le r \le 0.089$ respectively, which show excellent agreement with the recently released data \cite{pd1, pd2}. The number of e-foldings now varies within the acceptable range $46 < \mathrm{N} < 73$, which is sufficient to solve the horizon and flatness problems. For the sake of visualization we present the the spectral index of scalar perturbation versus the scalar to tensor ratio plot in figure 3. We  also present another pair of data sets in Table 3 and Table 4, for lower values of $\phi_i$. In both the tables, the spectral index of scalar perturbation and the scalar to tensor ratio lie within the range $0.969 \le n_s \le 0.979$ and $0.058\le r \le 0.083$ respectively, which again show excellent agreement with the recently released data \cite{pd1, pd2}. The number of e-folds ($\mathrm{N}$) also lie very much within admissible region (Table 3 and Table 4).\\

\begin{figure}

   \begin{minipage}[h]{0.47\textwidth}
      \centering
      \begin{tabular}{|c|c|c|c|c|}
      \hline\hline
      $\alpha_0$ in ${M_P}^3$ & $\phi_f$ in $M_P$ & $n_s$ & $r$ & $\mathrm{N}$\\
      \hline
      0.00019& 4.7077 & 0.9671 & 0.08883 & 46\\
      0.00018& 4.7079 & 0.9688 & 0.08415 & 48\\
      0.00017& 4.7080 & 0.9705 & 0.07948 & 51\\
      0.00016& 4.7082 & 0.9723 & 0.07480 & 55\\
      0.00015& 4.7084 & 0.9740 & 0.07013 & 58\\
      0.00014& 4.7086 & 0.9757 & 0.06545 & 62\\
      0.00013& 4.7088 & 0.9775 & 0.06078 &67\\
      0.00012& 4.7090 & 0.9792 & 0.05610 &73\\

       \hline\hline
    \end{tabular}
      \captionof{table}{Data set for the inflationary parameters taking $\phi_i=4.8 M_P$~;~$m^2=0.09{M_P}^2$;~$\Lambda=1.0 {M_P}^2$ and ~ varying $\alpha_0$.}
      \label{tab:table2}
   \end{minipage}%
   \hfill%
   \begin{minipage}[h]{0.47\textwidth}
      \centering
      \includegraphics[width=0.9\textwidth]{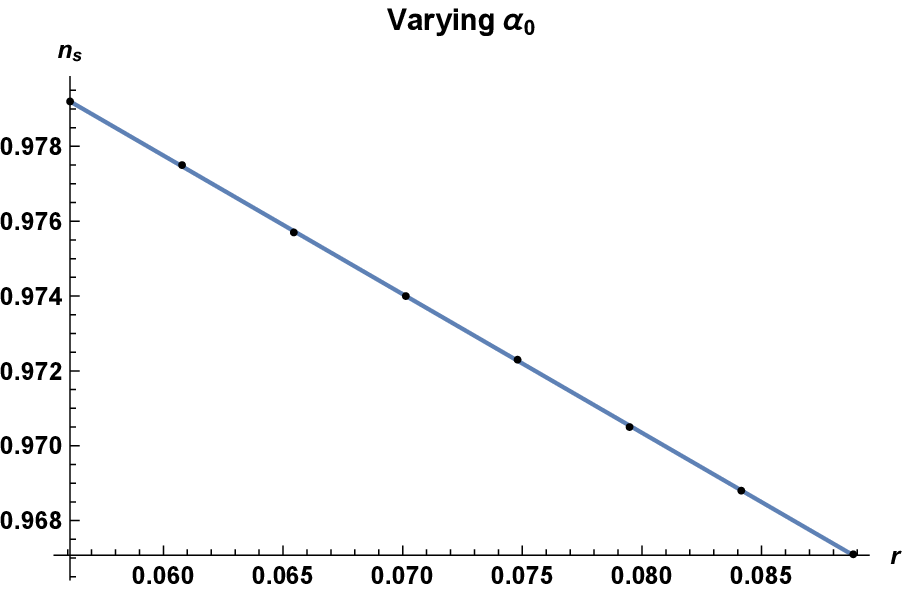}
      \caption{This plot depicts the variation of $n_s$ with $r$, varying $\alpha_0$.}
      \label{fig:fig3}
   \end{minipage}%

\end{figure}

\begin{figure}

   \begin{minipage}[h]{0.47\textwidth}
      \centering
      \begin{tabular}{|c|c|c|c|c|}
      \hline\hline
      $\alpha_0$ in ${M_P}^3$ & $\phi_f$ in $M_P$ & $n_s$ & $r$ & $\mathrm{N}$\\
      \hline
      0.000100& 1.9929 & 0.9696 & 0.08367 & 51\\
      0.000095& 1.9931 & 0.9711 & 0.07949 & 53\\
      0.000090& 1.9933 & 0.9726 & 0.07530 & 56\\
      0.000085& 1.9935 & 0.9741 & 0.07111 & 60\\
      0.000080& 1.9937 & 0.9756 & 0.06694 & 63\\
      0.000075& 1.9939 & 0.9772 & 0.06275 & 68\\
      0.000070& 1.9941 & 0.9787 & 0.05857 &72\\

       \hline\hline
    \end{tabular}
      \captionof{table}{Data set for the inflationary parameters taking $\phi_i=2.1 M_P$~;~$m^2=0.5{M_P}^2$;~$\Lambda=1.0 {M_P}^2$ and ~ varying $\alpha_0$.}
      \label{tab:table3}
   \end{minipage}%
\hfill%
\begin{minipage}[h]{0.47\textwidth}
      \centering
      \begin{tabular}{|c|c|c|c|c|}
      \hline\hline
      $\alpha_0$ in ${M_P}^3$ & $\phi_f$ in $M_P$ & $n_s$ & $r$ & $\mathrm{N}$\\
      \hline
      0.000095 & 1.4828 & 0.9673 & 0.09089 & 48\\
      0.000090 & 1.4830 & 0.9690 & 0.08610 & 50\\
      0.000085 & 1.4832 & 0.9708 & 0.08132 & 53\\
      0.000080 & 1.4834 & 0.9729 & 0.07654 & 57\\
      0.000075 & 1.4836 & 0.9742 & 0.07175 & 61\\
      0.000070 & 1.4839 & 0.9759 & 0.06697 & 65\\
      0.000065 & 1.4841 & 0.9776 & 0.06219 & 70\\
      0.000060 & 1.4844 & 0.9794 & 0.05740 &76\\

       \hline\hline
    \end{tabular}
      \captionof{table}{Data set for the inflationary parameters taking $\phi_i=1.6 M_P$~;~$m^2=0.9{M_P}^2$;~$\Lambda=1.0 {M_P}^2$ and ~ varying $\alpha_0$.}
      \label{tab:table4}
   \end{minipage}%
\end{figure}

Now, for a consistency check, let us chose sub-Planckian energy scale $\mathrm{H}_* \approx 10^{-5}M_P$, as in the previous subsection, together with a mid range value of $\alpha_0 =0.00015 M^3_P$, while $\Lambda = 1 M^2_P,~\phi_i=4.8 M_P$, are chosen as depicted in the table 2. Therefore, the term appearing in the left hand side of equation \eqref{HM} may be neglected, and we simply have,

\be{1\over 2}m^2\phi^2-\Lambda M^2_P \approx 0, \hspace{0.3in}\mathrm{resulting ~in},\hspace{0.3in}   m^2 \approx 0.09 M^2_P,\ee
which shows consistency with our data set presented in table 2. Further, since $r = 0.07$ for the above choice of $\alpha_0$, so we can calculate the energy scale of inflation $\mathrm{H_*}$ using the formula used in a single scalar field as \cite{Wands},
\be \mathrm{H_*}=8\times10^{13}{\sqrt{r\over 0.2}}GeV = 4.733 \times 10^{13} GeV \approx 1.97\times 10^{-5}M_P,\ee
which further reveals consistency in our choice of the energy scale $\mathrm{H}_*$ during inflation. From the analysis presented in the above two subsections, it is quite clear that inflation is supported by the vacuum energy density, rather than the dilaton, and thus the problem of moduli field stabilization does not appear.

\subsubsection{Inflation with different coupling parameters:}

Here, we exhibit that all the different choices of the coupling parameters $\alpha(\phi)$, $\beta(\phi)$ and $\gamma(\phi)$ presented in subcase 1b, and cases 2, 3 and 4 do not tell upon the inflationary parameters. \\

\noindent
\textbf{Subcase 1b:}\\

\noindent
Following equation \eqref{Mass}, here again we make the choice,

\be 12\mathrm{H}^2\left(\alpha+12\mathrm{H}^2\beta+2\mathrm{H}^2\gamma\right) \approx -\frac{\mathrm{H}^2\phi^2}{2}~~\mathrm{such~ that},~~ U = {1\over 2} {m^2\phi^2}-\Lambda M^2_P,~~\mathrm{where},~~ m^2 = \lambda^2+\mathrm{H}^2 \approx 2\lambda^2.\ee
Since the re-defined potential $U(\phi)$ together with the mass relation remain unaltered, so all the computations made above remains unaltered.\\

\noindent\textbf{Case 2:}\\

\noindent
Even under the choice of $\beta= \beta_0 =$ constant made in case 2 \eqref{betagamma2}, we obtain the re-defined potential $U(\phi)$ in view of \eqref{Mass} as,

\be 12\mathrm{H}^2\left(\alpha+2\mathrm{H}^2\gamma\right) \approx -\frac{\mathrm{H}^2\phi^2}{2}~~\mathrm{such~ that},~~ U = {1\over 2} {m^2\phi^2}-\Lambda M^2_P,~~\mathrm{where},~~ m^2 = \lambda^2+\mathrm{H}^2 \approx 2\lambda^2.\ee
Thus, $U(\phi)$ together with the mass formula remain unaltered here again, and hence as mentioned all the computations remain valid.\\

\noindent
\textbf{Case 3:}\\

\noindent
Again the choice of $\alpha= \alpha_0 =$ constant made in case 3 in \eqref{betagamma3} does not affect the re-defined potential $U(\phi)$ and the mass formula since,

\be 12\mathrm{H}^2\left(12\mathrm{H}^2\beta+2\mathrm{H}^2\gamma\right) \approx -\frac{\mathrm{H}^2\phi^2}{2}~~\mathrm{such~ that},~~ U = {1\over 2} {m^2\phi^2}-\Lambda M^2_P +6\lambda^2\alpha_0,~~\mathrm{where},~~ m^2 = \lambda^2+\mathrm{H}^2 \approx 2\lambda^2.\ee

\noindent
\textbf{Case 4:}\\

\noindent
Finally, keeping both the coupling parameters $\alpha= \alpha_0=$ constant and $\beta= \beta_0 =$ constant, made in case 4 \eqref{betagamma4} one can find the form of the re-defined potential $U(\phi)$ as,

\be 12\mathrm{H}^2\left(2\mathrm{H}^2\gamma\right) \approx -\frac{\mathrm{H}^2\phi^2}{2}~~\mathrm{such~ that},~~ U = {1\over 2} {m^2\phi^2}-\Lambda M^2_P + 6\lambda^2\alpha_0,~~\mathrm{where},~~ m^2 = \lambda^2+\mathrm{H}^2 \approx 2\lambda^2,\ee
which is the same as before. Note that the additional constant term ($6\lambda^2\alpha_0$) appearing in the potential of cases 3 and 4, is extremely small to make any difference This also validates the choice of the action \eqref{A}, we begin with.

\subsubsection{Graceful exit:}

Although the theory under consideration is highly complicated, we have been able to reduce the system of field equations considerably to study inflation. We now pose to prove that the theory does not also suffer from graceful exit problem. For this purpose, we need to show that as $\phi \ll M_P$, it executes oscillatory behaviour. Thus, at first we have to prove that $\phi$ indeed falls of with time. Since both $\lambda$ as well as $\alpha_0$ are about five order of magnitude smaller than the other parameters, we therefore express equation \eqref{Consistency} under suitable approximation as:

\be  \dot\phi \approx -{M_P^2\Lambda\over {\lambda\phi}} ,\ee
which may immediately be solved (taking $\lambda \approx 10^{-5} M_P,~\Lambda =1 M^2_P$) as,
\be\left\{\phi (t)=-\sqrt{2} \sqrt{c_1-100000 t} M_P\right\}~~\mathrm{or}~~\left\{\phi (t)=\sqrt{2} \sqrt{c_1-100000 t}  M_P\right\}.\ee
\begin{figure}
\begin{minipage}[h]{0.47\textwidth}
      \centering
      \includegraphics[width=0.9\textwidth]{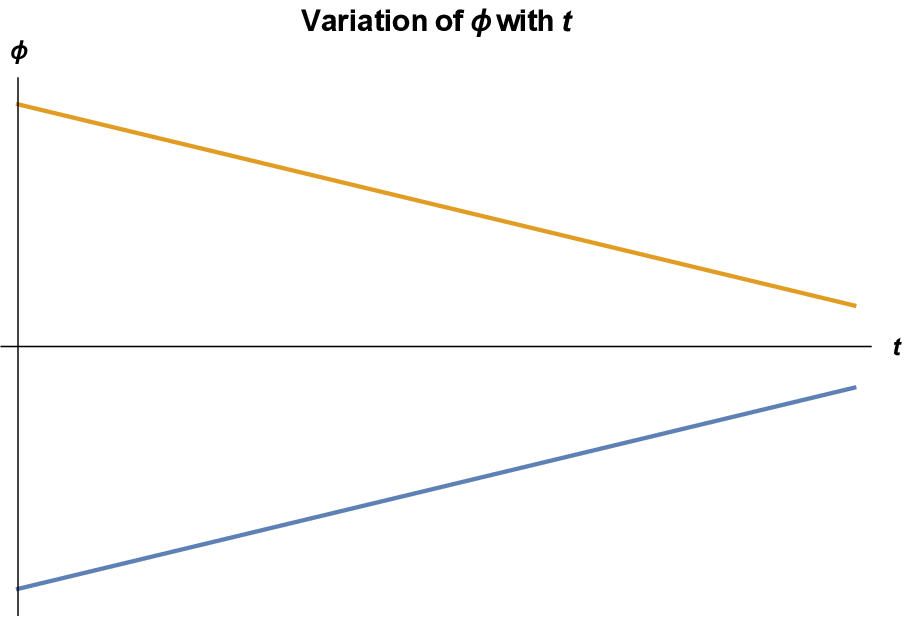}
      \caption{This plot depicts fall of $\phi$ with $t$.}
      \label{fig:fig4}
   \end{minipage}%

\end{figure}

\noindent
We present Figure 4 to depict the behaviour of $\phi$ with time, taking the integration constant, $c_1 = 3$. Having shown that $\phi$ decays, let us now express equation \eqref{Fried} as,

\be 3\mathrm{H}^2 = {1\over 2\alpha}\left({1\over 2}\dot\phi^2 + {1\over 2} m^2 \phi^2-\Lambda M^2_P\right),\ee
taking, $U(\phi) = {1\over 2} m^2 \phi^2-\Lambda M^2_P$, and $m^2 \approx 2 \lambda^2$. In view of the expression of $\alpha(\phi) = {\alpha_0\over \phi}$, the above equation reads as,

\be \label{Hubble} {3\mathrm{H}^2\over m^2} = {3\mathrm{H}^2\over 2\lambda^2} = {\phi\over 4\alpha_0}\left({\dot\phi^2\over 2\lambda^2} + \phi^2-{\Lambda M^2_P\over 2\lambda^2}\right).\ee
Note that for single scalar field, the above equation reads as $3\mathrm{H}^2 = {1\over 2M_p^2}(\dot\phi^2 + m^2 \phi^2-2\Lambda M^2_P)$. Since at the end of inflation, ${\phi\over 4\alpha_0} \sim {7359\over M_p^2}$, according to the present data set, so once the Hubble rate falls below $\sqrt {2}\lambda$, this equation \eqref{Hubble} may be approximated to,

\be \dot\phi^2 \approx -(2\lambda^2 \phi^2-\Lambda M^2_P),\ee
which may immediately be integrated to yield,

\be\phi (t)=\pm\frac{\sqrt{\Lambda }M_p \tan \left(\sqrt{2} \lambda \{t- t_0\} \right)}{\lambda  \sqrt{2[\tan ^2\left(\sqrt{2} \lambda \{t- t_0\} \right)+1]}} = \pm \frac{\sqrt{\Lambda}M_p }{\sqrt{2}\lambda} \sin \left(\sqrt{2} \lambda \{t- t_0\} \right).\ee
$t_0$ being a constant of integration. Thus the scalar field (dilaton) starts oscillating many times over a Hubble time, driving a matter-dominated era at the end of inflation.

\section{\bf{Concluding remarks}}

Although, there are alternatives to inflation, such as matter bounce or ekpyrosis originating from string theory, branes and extra dimensions \cite{50,51}, which do explain the origin of large scale structure and flatness of the universe \cite{52,53}, inflation is indeed the mainstream choice. Inflation perhaps, is the simplest scenario, which not only solves the horizon and the flatness problems but also can explain elegantly, the origin of the seeds of perturbation. It is therefore required to test every alternative/modified theory of gravity in the context of inflation. As mentioned, inflation is a quantum theory of perturbation, and it occurred when gravity becomes classical, i.e., at the sub-Planckian epoch. It is therefore also required to test the viability of the alternative/modified theory models under quantization. In the present manuscript, a generalized action has been constructed out of non-minimally coupled [$\alpha(\phi)$] scalar-tensor theory of gravity, being associated with a curvature squared term having functionally dependent coupling parameter [$\beta(\phi)$], the Gauss-Bonnet-Dilatonic [$\gamma(\phi)$] term and the cosmological constant term. Such an action has not been treated in the context of inflation, to the best of our knowledge. A variety of classical de-Sitter solutions have been presented, which exhibit identical form of the potential under the choice of different coupling parameters, and also with constant coupling parameters $\alpha(\phi) = \alpha_0$, or $\beta(\phi) = \beta_0$, or both. This validates the choice of a complicated action to begin with. The phase-space structure has been formulated following `MHF'. The effective Hamiltonian under quantization has been found to be hermitian, establishing unitarity, also a viable semiclassical approximation has been presented which depicts that the quantum equation is classically allowed, and the semiclassical wave function is strongly peaked around a classical de-Sitter solution. This motivates to study inflation.\\

One important finding is that, in the presence of dilatonic coupling, a combined hierarchy of Hubble and Gauss–Bonnet flow parameters are usually required to handle the situation, since additional condition apart from the standard slow roll condition are needed \cite{We2, GB2}. Also, in the case of a non-minimally coupled scalar-tensor theory of gravity in the presence of scalar curvature squared term with constant coupling (without dilaton) again a combined hierarchy of Hubble and non-minimal flow parameters was introduced \cite{jhep}. It therefore appears that additional conditions are essential corresponding to the number of coupling parameters present in the theory. However, although in the present case, there are three independent functional coupling parameters, we do not require any further condition, like the hierarchy of flow parameters associated with non-minimal coupling parameter $\alpha(\phi)$, $\beta(\phi)$ coupled with $R^2$ and $\gamma(\phi)$: the dilatonic Gauss–Bonnet coupling. In fact, a combined hierarchy of Hubble flow parameter has been found to be sufficient to study inflation, while all the information regarding the coupling parameters is carried by the re-defined effective potential $U(\phi)$. The choice of the quadratic potential, and the functional forms of coupling parameters are not arbitrary, rather they satisfy classical de-Sitter solution.\\

In the absence of the cosmological constant term, although the inflationary parameters (scalar to tensor ratio, $r$ and the spectral index, $n_s$) obtained show reasonably good agreement with the latest released Planck's data \cite{pd1, pd2}, nevertheless, the number of e-fold is too large, which might cause problem in reheating the universe. This problem is due to the fact that the same dilaton has been called for inflation, and it is well known that dilaton suffers from stabilization issue. In the presence of the cosmological constant term, the problem disappears and the fit is excellent including the number of e-folds. Since the numerical value of the cosmological constant has been chosen to be $\Lambda = 1.0 ~M_P^2$, which corresponds to the energy density $\rho_\Lambda \approx 10 ^{73}~GeV^4$, which is the computed value of the vacuum energy density, $\rho_{\mathrm{vac}}$, it is therefore apparent that inflation is essentially driven by the vacuum energy density, and the dialton stabilization issue is bypassed. Thus the scalar field may be treated as the moduli field (dilaton), rather than an ordinary scalar field.\\

Last but not the least, it has been expatiated that all the different choices of the coupling parameters which lead to the same classical de-Sitter solution with identical potential, lead to the same re-defined effective potential [$U(\phi)$], and as a result inflationary parameters remain unaltered. This not only validates the choice of the complicated action we begin with, but also the technique of using effective potential instead of taking non-minimal flow parameters into account.

\appendix

\section{Canonical formulation following Dirac's algorithm:}

In Dirac's constraint analysis, it is customary to fix $h_{ij}$ and $K_{ij}$ at the boundary, and therefore there is no need to supplement the action by surface terms. Hence, one can consider the action (\ref{Sc1}) without the supplementary boundary terms and integrate the divergent terms appearing in the action by parts, may express the point Lagrangian in the form,
\begin{center}\be \label{A1}\begin{split} L&= {\bigg{(}-\frac{3\alpha'\dot\phi\dot z\sqrt z}{N}-\frac{3\alpha\dot z^2}{2N\sqrt z}+6kN\alpha\sqrt z\bigg{)}}-Nz^{3\over 2}\Lambda M_p^2+\frac{9\beta}{\sqrt z}\bigg{(}\frac{\ddot z^2}{N^3}-\frac{2\dot z \ddot z \dot N}{N^4}+\frac{\dot z^2\dot N^2}{N^5}+\frac{2k{\dot z}^2}{Nz}+4k^2N \bigg{)}\\&-\frac{36\beta' k \dot z \dot\phi}{N\sqrt z}-\frac{\gamma'\dot z \dot\phi}{N\sqrt z}\bigg{(}\frac{\dot z^2}{N^2z}+12k \bigg{)}+z^{\frac{3}{2}}\bigg{(}\frac{1}{2N}\dot{\phi}^2-VN\bigg{)}, \end{split}\ee\end{center}
which is essentially the action (\ref{Sc2}) without the boundary term $\Sigma_{R_2^2}$. Thereafter, it is required to substitute $\dot z=Nx$, i.e.; $\ddot z=N\dot x+\dot N x$, so that the point Lagrangian \eqref{A1} may be express in the following form,

\begin{center}\be \label{A2}\begin{split} L&= {\bigg{(}-{3\alpha'\dot\phi \sqrt z x}-\frac{3\alpha N x^2}{2\sqrt z}+6kN\alpha\sqrt z\bigg{)}}-Nz^{3\over 2}\Lambda M_p^2+\frac{9\beta}{\sqrt z}\bigg{(}\frac{\dot x^2}{N}+\frac{2kN x^2}{z}+4k^2N \bigg{)}-\frac{36\beta' k x \dot\phi}{\sqrt z}-\frac{\gamma'x \dot\phi}{\sqrt z}\bigg{(}\frac{x^2}{z}+12k \bigg{)}\\& +z^{\frac{3}{2}} \bigg{(}\frac{1}{2N}\dot{\phi}^2-VN\bigg{)}+u\bigg({\dot z\over N}-x \bigg), \end{split}\ee\end{center}
where the expression $\big({\dot z\over N}-x \big)$ is being treated as a constraint and introduced through the Lagrangian multiplier $u$.The canonical momenta are,

\begin{eqnarray}
  p_x &=& {18\beta \dot x\over N\sqrt z}; ~~ p_z={u\over N};~~p_{\phi}=-3\alpha'\sqrt z x-{36\beta' kx\over \sqrt z}-{\gamma' x\over \sqrt z}{\big({x^2\over z}+12k\big)}+{\dot\phi z^{3\over 2}\over N};~~p_N=0=p_u.
\end{eqnarray}
The primary Hamiltonian is Therefore,

\begin{equation}\label{AHp}\begin{split}
   & H_{p_1}=3\alpha \bigg{(}\frac{Nx^2}{2\sqrt z}-2kN\sqrt z\bigg{)}+{u\dot z\over N}+\frac{3\alpha'N x p_{\phi}}{z}+\frac{9\alpha'^2 Nx^2}{2\sqrt z} + \frac{\sqrt z N p_x^2}{36\beta}-\frac{18k\beta N}{\sqrt z}\bigg{(}\frac{x^2}{z}+2k\bigg{)}+\frac{\gamma'^2N {x}^6}{2z^{\frac{9}{2}}}\\& +\frac{648k^2N\beta'^2{x}^2}{z^{\frac{5}{2}}} + \frac{72Nk^2\gamma'^2 x^2}{z^{\frac{5}{2}}} +\frac{36k\beta'{Nx} {p_{\phi}}}{z^2}+\frac{12kN\gamma'^2 {x}^4}{z^{\frac{7}{2}}}+\frac{\gamma' {N{x}^3}{p_{\phi}}}{z^3} +\frac{12k\gamma' Nx p_{\phi}}{z^2} +\frac{36k\beta'\gamma' Nx^4}{z^{\frac{7}{2}}}\\& +\frac{432k^2\beta'\gamma' Nx^2}{z^{\frac{5}{2}}} +\frac{Np_{\phi}^2}{2z^{\frac{3}{2}}}+\frac{108k\alpha'\beta' Nx^2}{z^{\frac{3}{2}}} +\frac{3\alpha' \gamma' Nx^4}{z^{\frac{5}{2}}}+\frac{36k\alpha'\gamma' Nx^2}{z^{\frac{3}{2}}} +VNz^{\frac{3}{2}}+ Nz^{3\over 2}\Lambda M_p^2-u\big( {\dot z\over N}-x\big).
\end{split}
\end{equation}
Now introducing the constraints $\phi_1= Np_z-u  \approx 0$ and $\phi_2=p_u \approx 0$ through the Lagrange multipliers $u_1$ and $u_2$ respectively, we get (Note that, these are second class constraints \footnote{Although, second class constraints are handled with Dirac brackets, one can trivially check that the Poisson bracket is identical to the Dirac bracket, in the present situation.}, since $\{\phi_i, \phi_j\} \ne 0$. Further, since the lapse function $N$ is non-dynamical, so the associated constraint vanishes strongly, ie. $p_N =0$ and therefore it may safely be ignored),
\begin{equation}\label{AHp1}\begin{split}
    H_{p_1}&=N\bigg[3\alpha \bigg{(}\frac{x^2}{2\sqrt z}-2k\sqrt z\bigg{)}+\frac{3\alpha' x p_{\phi}}{z}+\frac{9\alpha'^2 x^2}{2\sqrt z} + \frac{\sqrt z  p_x^2}{36\beta}-\frac{18k\beta }{\sqrt z}\bigg{(}\frac{x^2}{z}+2k\bigg{)}+\frac{\gamma'^2 {x}^6}{2z^{\frac{9}{2}}}\\& +\frac{648k^2\beta'^2{x}^2}{z^{\frac{5}{2}}} + \frac{72k^2\gamma'^2 x^2}{z^{\frac{5}{2}}} +\frac{36k\beta'{x} {p_{\phi}}}{z^2}+\frac{12k\gamma'^2 {x}^4}{z^{\frac{7}{2}}}+\frac{\gamma' {{x}^3}{p_{\phi}}}{z^3} +\frac{12k\gamma' x p_{\phi}}{z^2} +\frac{36k\beta'\gamma' x^4}{z^{\frac{7}{2}}}\\& +\frac{432k^2\beta'\gamma' x^2}{z^{\frac{5}{2}}} +\frac{p_{\phi}^2}{2z^{\frac{3}{2}}}+\frac{108k\alpha'\beta' x^2}{z^{\frac{3}{2}}} +\frac{3\alpha' \gamma' x^4}{z^{\frac{5}{2}}}+\frac{36k\alpha'\gamma' x^2}{z^{\frac{3}{2}}} +Vz^{\frac{3}{2}}+z^{3\over 2}\Lambda M_p^2\bigg]+ux+u_1\big(Np_z-u\big)+u_2p_u.
\end{split}
\end{equation}
Note that the Poisson brackets $\{x, p_x\} = \{z, p_z\} = \{\phi,p_{\phi}\}=\{u, p_u\} = 1$, hold. Now constraint should remain preserved in time, which are exhibited through the following Poisson brackets
\begin{equation}\label{Apc}
   \dot\phi_1=\{ \phi_1,H_{p_1}\}=-u_2-N{\partial H_{p_1}\over \partial z}\approx 0\Rightarrow u_2=-N{{\partial H_{p_1}\over \partial z}};~ \dot\phi_2= \{\phi_2,H_{p_1}\}\approx 0 = u_1 - x \Rightarrow u_1=x.
\end{equation}
Therefore the primary Hamiltonian is modified to
\begin{equation}\label{AHp2}\begin{split}
    H_{p_2}&=N\bigg[xp_z+3\alpha \bigg{(}\frac{x^2}{2\sqrt z}-2k\sqrt z\bigg{)}+\frac{3\alpha' x p_{\phi}}{z}+\frac{9\alpha'^2 x^2}{2\sqrt z} + \frac{\sqrt z  p_x^2}{36\beta}-\frac{18k\beta }{\sqrt z}\bigg{(}\frac{x^2}{z}+2k\bigg{)}+\frac{\gamma'^2 {x}^6}{2z^{\frac{9}{2}}}\\& +\frac{648k^2\beta'^2{x}^2}{z^{\frac{5}{2}}} + \frac{72k^2\gamma'^2 x^2}{z^{\frac{5}{2}}} +\frac{36k\beta'{x} {p_{\phi}}}{z^2}+\frac{12k\gamma'^2 {x}^4}{z^{\frac{7}{2}}}+\frac{\gamma' {{x}^3}{p_{\phi}}}{z^3} +\frac{12k\gamma' x p_{\phi}}{z^2} +\frac{36k\beta'\gamma' x^4}{z^{\frac{7}{2}}}\\& +\frac{432k^2\beta'\gamma' x^2}{z^{\frac{5}{2}}} +\frac{p_{\phi}^2}{2z^{\frac{3}{2}}}+\frac{108k\alpha'\beta' x^2}{z^{\frac{3}{2}}} +\frac{3\alpha' \gamma' x^4}{z^{\frac{5}{2}}}+\frac{36k\alpha'\gamma' x^2}{z^{\frac{3}{2}}} +Vz^{\frac{3}{2}}+z^{3\over 2}\Lambda M_p^2\bigg]-Np_u{\partial H_{p_1}\over \partial z}.
\end{split}
\end{equation}
As the constraint should remain preserved in time in the sense of Dirac, so
\begin{equation}\label{Hp2}
    \dot\phi_1=\{\phi_1,H_{p2}\}=-N\bigg[{\partial H_{p_1}\over \partial z}-Np_u{\partial^2H_{p_1}\over \partial z^2} \bigg]+N{\partial H_{p_1}\over \partial z}
\approx 0\Rightarrow p_u=0.
\end{equation}
Finally the phase-space structure of the Hamiltonian, being free from constraints reads as,
\begin{equation}\label{AHF}\begin{split}
    H&=N\bigg[xp_z+3\alpha \bigg{(}\frac{x^2}{2\sqrt z}-2k\sqrt z\bigg{)}+\frac{3\alpha' x p_{\phi}}{z}+\frac{9\alpha'^2 x^2}{2\sqrt z} + \frac{\sqrt z  p_x^2}{36\beta}-\frac{18k\beta }{\sqrt z}\bigg{(}\frac{x^2}{z}+2k\bigg{)}+\frac{\gamma'^2 {x}^6}{2z^{\frac{9}{2}}}\\& +\frac{648k^2\beta'^2{x}^2}{z^{\frac{5}{2}}} + \frac{72k^2\gamma'^2 x^2}{z^{\frac{5}{2}}} +\frac{36k\beta'{x} {p_{\phi}}}{z^2}+\frac{12k\gamma'^2 {x}^4}{z^{\frac{7}{2}}}+\frac{\gamma' {{x}^3}{p_{\phi}}}{z^3} +\frac{12k\gamma' x p_{\phi}}{z^2} +\frac{36k\beta'\gamma' x^4}{z^{\frac{7}{2}}}\\& +\frac{432k^2\beta'\gamma' x^2}{z^{\frac{5}{2}}} +\frac{p_{\phi}^2}{2z^{\frac{3}{2}}}+\frac{108k\alpha'\beta' x^2}{z^{\frac{3}{2}}} +\frac{3\alpha' \gamma' x^4}{z^{\frac{5}{2}}}+\frac{36k\alpha'\gamma' x^2}{z^{\frac{3}{2}}} +Vz^{\frac{3}{2}}+z^{3\over 2}\Lambda M_p^2\bigg]=N\mathcal{H},
\end{split}
\end{equation}
which is exactly the same as (\ref{Hc}). This establishes the fact that Dirac formalism after taking care of the divergent terms, is identical to the Modified Horowitz' formalism.

\section{Field equations from Hamilton's equations:}

In this appendix, we pose to find all the field equations \eqref{avariation}, \eqref{00}, and \eqref{phivariation}, obtained in the Robertson-Walker minisuperspace \eqref{RW} under the variation of the action \eqref{phivar1} from the Hamiltonian \eqref{Hc}. We therefore find the Hamilton's equations and combine to form the Euler-Lagrange equations. The Hamilton's equations are,

\be \label{xdot} \dot{x}=\frac{\partial H}{\partial p_x}=\frac{\sqrt{z}p_x}{18\beta} \ee

\be \label{phidot}  \dot{\phi}=\frac{\partial H}{\partial p_{\phi}}=\frac{3\alpha' x}{z}+\frac{36k\beta'x}{z^2}+\frac{\gamma'x}{z^2}\bigg(\frac{x^2}{z}+12k\bigg)+\frac{p_{\phi}}{z^\frac{3}{2}} \ee

\be \label{pphi} \begin{split} \dot{p_{\phi}}=-\frac{\partial H}{\partial \phi} = -&\bigg{[}3\alpha' \bigg{(}\frac{x^2}{2\sqrt z}-2k\sqrt z\bigg{)}- \frac{\beta'\sqrt z p_x^2}{36\beta^2}+\frac{3\alpha'' x p_{\phi}}{z}+\frac{36k\beta''{x} {p_{\phi}}}{z^2}+\frac{\gamma'' xp_{\phi}}{z^2}\bigg(\frac{x^2}{z}+12k\bigg)\\&+\frac{9\alpha'\alpha'' x^2}{\sqrt z}-\frac{18k\beta'}{\sqrt z}\bigg{(}\frac{x^2}{z}+2k\bigg{)}+\frac{1296k^2\beta'\beta''{x}^2}{z^{\frac{5}{2}}}+\frac{108k\alpha''\beta' x^2}{z^{\frac{3}{2}}}+\frac{108k\alpha'\beta'' x^2}{z^{\frac{3}{2}}}\\&+\frac{3\alpha''\gamma'x^2}{z^\frac{3}{2}}\bigg(\frac{x^2}{z}+12k\bigg)+\frac{3\alpha'\gamma''x^2}{z^\frac{3}{2}}\bigg(\frac{x^2}{z}
+12k\bigg)+\frac{36k\beta''\gamma'x^2}{z^\frac{5}{2}}\bigg(\frac{x^2}{z}+12k\bigg)\\&+\frac{36k\beta'\gamma''x^2}{z^\frac{5}{2}}\bigg(\frac{x^2}{z}
+12k\bigg)+\frac{\gamma'\gamma''x^2}{z^\frac{5}{2}}\bigg(\frac{x^2}{z}+12k\bigg)^2+V'z^{\frac{3}{2}}\bigg{]}. \end{split}\ee
Now inserting $p_x$ and $p_{\phi}$ in the above equation \eqref{pphi} from equations \eqref{xdot} and \eqref{phidot} respectively, one can immediately reduced equation \eqref{pphi} as

\be\label{pphidot}\begin{split}\dot{p_{\phi}}=\bigg{[}-3\alpha' \bigg{(}\frac{x^2}{2\sqrt z}-2k\sqrt z\bigg{)}-3\alpha''x\sqrt{z}\dot{\phi}+\frac{9\beta'\dot{x}^2}{\sqrt{z}}+\frac{18k\beta'}{\sqrt z}\bigg{(}\frac{x^2}{z}+2k\bigg{)}-\frac{\gamma'' x\dot{\phi}}{\sqrt{z}}\bigg(\frac{x^2}{z}+12k\bigg)-\frac{36k\beta''x\dot{\phi}}{\sqrt{z}}-V'z^\frac{3}{2}\bigg{]},\end{split}\ee
performing temporal derivative on equation \eqref{phidot}, we obtained

\be\label{pphidot2}\begin{split}\dot{p_{\phi}}=\bigg{[}&\ddot{\phi}z^\frac{3}{2}+\frac{3\sqrt{z}\dot{z}\dot{\phi}}{2}-3\alpha''x\sqrt{z}\dot{\phi}
-3\alpha'\dot{x}\sqrt{z}-\frac{3\alpha'x\dot{z}}{2\sqrt{z}}-\frac{36k\beta''x\dot{\phi}}{\sqrt{z}}-\frac{36k\beta'\dot{x}}{\sqrt{z}}+\frac{18k\beta'x\dot{z}}
{z^\frac{3}{2}}-\frac{\gamma''x\dot{\phi}}{\sqrt{z}}\bigg(\frac{x^2}{z}+12k\bigg)\\&-\frac{\gamma'\dot{x}}{\sqrt{z}}\bigg(\frac{3x^2}{z}+12k\bigg)+\frac{\gamma' x\dot{z}}{2z^\frac{3}{2}}\bigg(\frac{3x^2}{z}+12k\bigg)\bigg{]}.\end{split}\ee
Therefore, comparing \eqref{pphidot} and \eqref{pphidot2} one can readily arrive at,

\be \label{phivariation1} \begin{split}& -6\alpha'\bigg{(}a^2\ddot a+a\dot a^2+ka\bigg{)} -36\beta'\bigg{(}a\ddot a^2+2\dot a^2\ddot a+ \frac{\dot a^4}{a}+{k^2\over a}+{2k\dot a^2\over a}+2k\ddot a\bigg{)} +3a^2\dot a \dot\phi -24\gamma'\bigg{(}\dot a^2\ddot a+k\ddot a\bigg{)} \\&+a^3\bigg{(}\ddot\phi+V'\bigg{)}=0.\end{split}\ee
If we now insert all the momenta in the Hamiltonian constraint equation \eqref{Hc}, the $(^0_0)$ equation of Einstein is obtained as,

\be \label{001} \begin{split} & -\frac{6\alpha}{a^2}\bigg{(}\dot a^2+ k\bigg{)}-\frac{6\alpha' \dot a\dot\phi}{a}+\Lambda M_p^2 -36\beta \bigg{(}\frac{2\dot a\dddot a}{a^2}-\frac{\ddot a^2}{a^2} +\frac{2\dot a^2 \ddot a}{a^3}-\frac{3\dot a^4}{a^4}-{2k\dot a^2\over a^4}+{k^2\over a^4}\bigg{)}-72\beta'\dot\phi\bigg{(}\frac{\dot a\ddot a}{a^2}+\frac{\dot a^3}{a^3}+{k\dot a\over a^3}\bigg{)}\\&-24\gamma'\dot\phi\bigg{(}{\dot a^3\over a^3}+{k\dot a\over a^3}\bigg{)}+\bigg{(}\frac{\dot\phi^2}{2}+V \bigg{)}=0. \end{split} \ee
\eqref{avariation} equation can also be obtained from the Hamilton's equation, which is a bit more complicated. Nevertheless, one can better combine the above equations \eqref{phivariation} and \eqref{00}, to obtain it in a straight forward manner. This establishes that the Hamiltonian obtained following Modified Horowitz' formalism  or equivalently following Dirac's constrained analysis \eqref{AHF} gives the correct description of the system under consideration.

\end{document}